# Experimentally Grounded Machine Learning of Precursor-Surface-Reactant Chemistry in Atomic Layer Deposition

Bonwook Gu[1], Trinh Ngoc Le[2], Zunair Masroor[1], Minhyeok Lee[1] and Han-Bo-Ram Lee[2*]

[1]Department of Materials Science and Engineering, Incheon National University, Incheon 22012, Korea

[2]Graduate School of Semiconductor Materials and Devices Engineering, Ulsan National Institute of Science and Technology (UNIST), Ulsan, 44919, Korea

***Corresponding Author**: hbrlee@unist.ac.kr

**Abstract**

Whether a precursor–substrate–reactant combination yields an atomic layer deposition (ALD) process is still settled by trial-and-error experiments. Here we introduce AJIN (ALD Joint Interpretable Network), a machine-learning framework that learns this decision from experimental outcomes. We compiled 560 experimentally observed ALD and area-selective ALD outcomes spanning 76 precursors, 27 substrates, 20 reactants and 25–620 °C. Following the ALD cycle, AJIN predicts adsorption from graph-neural-network embeddings of the precursor and substrate with temperature, and deposition by adding the reactant. The best model (CHGNet-MLP) reaches balanced accuracies of 79.2% for adsorption and 84.8% for deposition and transfers partially to unseen precursors. SHAP and ablation analyses show that substrate chemistry governs adsorption, whereas precursor, substrate and reactant all contribute to deposition. Tested against our laboratory's ALD experiments, AJIN correctly predicted 83% of new-substrate cases and 71% of new precursor–substrate adsorption cases, providing an experimentally grounded tool for prioritizing ALD and area-selective ALD chemistries.

## 1. Introduction

Atomic layer deposition (ALD) is a self-limiting thin-film deposition technique that achieves angstrom-level thickness control through sequential, alternating exposures of a precursor gas and a reactant gas to a substrate surface[1]. Owing to its exceptional conformality and atomic-scale precision, ALD has become indispensable in advanced semiconductor manufacturing, including gate dielectric deposition, diffusion barrier formation, and nanoscale patterning[1 2 3]. Each ALD cycle proceeds through sequential precursor and reactant exposures. During the first half-cycle, the precursor adsorbs onto the substrate surface. After the subsequent reactant exposure, deposition is determined by whether the complete precursor/reactant cycle produces measurable film growth. Therefore, adsorption and deposition are sequentially related outcomes governed by the specific combination of precursor, substrate, reactant, and process temperature.

Being able to predict whether a given precursor will adsorb or not adsorb on a target substrate, and whether subsequent reactant exposure will lead to deposition or non-deposition, would directly inform two central tasks in ALD process development. These tasks are prioritizing which candidate precursors are worth synthesizing and identifying which precursor–substrate–reactant–temperature combinations are likely to yield a viable ALD process. Beyond simply assigning binary outcomes, such a predictive framework can also help reveal which structural and process factors are responsible for a given result, offering mechanistic insight into why a particular precursor–substrate–reactant combination leads to adsorption, non-adsorption, deposition, or non-deposition. This predictive and explanatory capability is therefore valuable not merely as a description of ALD chemistry, but as a practical tool for guiding precursor design and process screening before committing to experimental synthesis and testing.

Experimentally screening the vast combinatorial space of precursors, substrates, reactants, and process temperatures for ALD process development is already prohibitively resource-intensive[4 5 6]. The number of candidate ALD systems grows multiplicatively with the diversity of each component: with numerous precursor chemistries, substrate materials, and reactant options spanning a wide temperature window (typically 25–620 °C), exhaustive experimental coverage is practically unattainable. This experimental bottleneck motivates a data-driven approach: if researchers can predict the outcomes of adsorption and deposition from structural and process information, the experimental search space can be dramatically narrowed, accelerating the discovery of viable ALD chemistries.

Previous computational and data-driven efforts to guide ALD process design have taken several forms. Density functional theory (DFT) calculations have been widely employed to evaluate adsorption energies and reaction pathways for specific precursor-surface pairs, providing atomistic insight into selectivity mechanisms[7 8 9 10 11]. However, DFT-based screening requires expert selection of candidate systems and significant computational resources, making it impractical for broad combinatorial exploration. More recently, DFT data have been coupled with machine learning (ML) models to extrapolate reactivity predictions across a larger chemical space[12 13 14]; yet such hybrid workflows inherit the same constraint that the practitioner must first curate a targeted set of systems for DFT calculation, and results vary substantially depending on which systems are selected. ML models trained on one-dimensional (1D) molecular descriptors have also been applied to ALD-related property prediction[5 6], but these representations compress structural information into predefined scalar descriptors rather than directly learning from three-dimensional atomistic graphs. Beyond this representational limitation, the sequential relationship between precursor adsorption and full-cycle deposition outcomes has received comparatively little attention in data-driven ALD prediction. In most existing approaches, adsorption- and deposition-related outcomes are treated as separate prediction problems with independently constructed feature sets, rather than as sequentially related outcomes within a consistent feature pipeline.

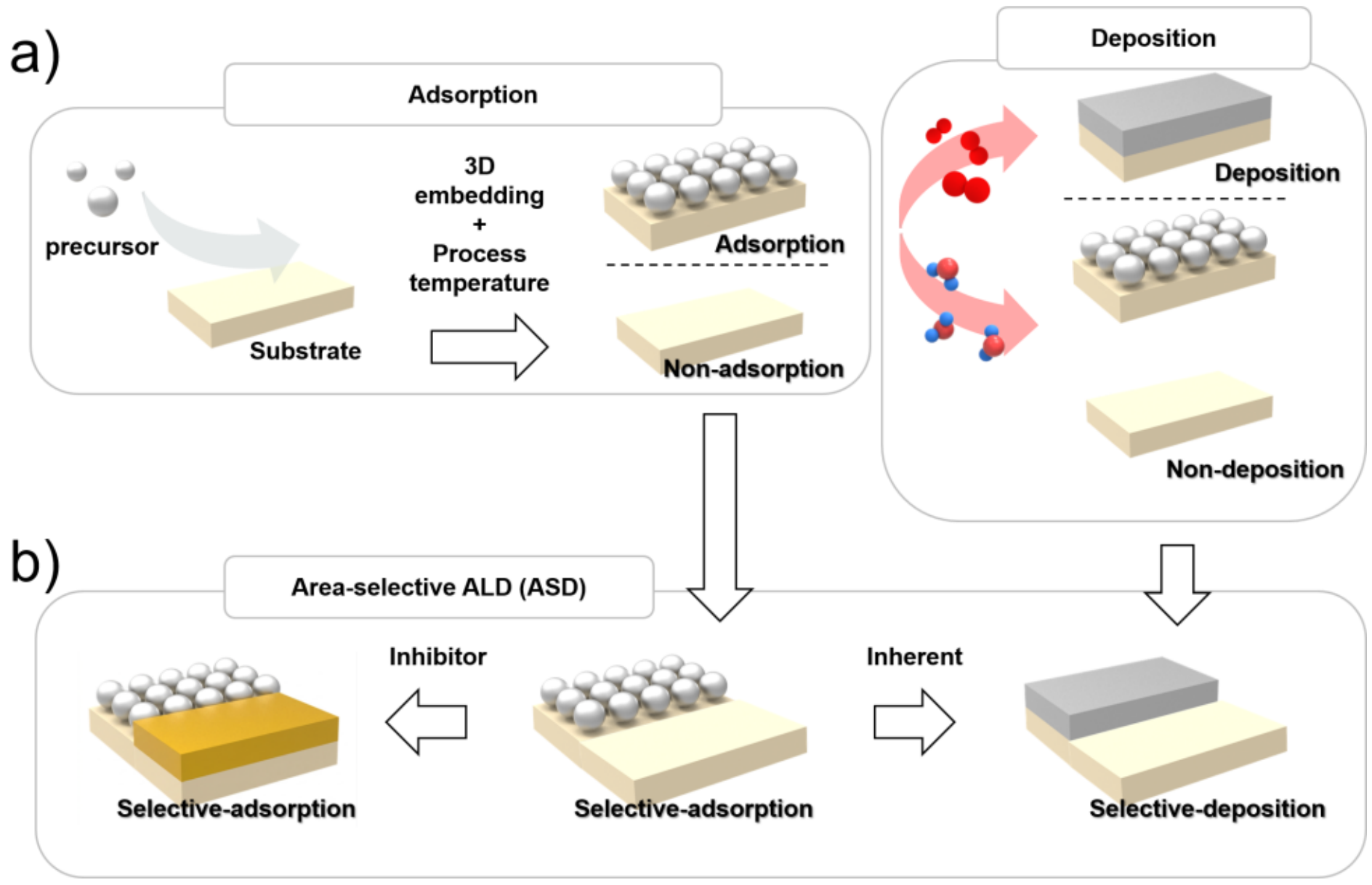

Figure 1. (a) Adsorption is formulated as a binary classification task using three-dimensional structural embeddings of the precursor and substrate together with process temperature, yielding adsorption or non-adsorption. Following the sequential chemistry of ALD, the same framework is extended to deposition by additionally incorporating the structural embedding of the reactant, yielding deposition or non-deposition for the given precursor–substrate–reactant–temperature combination. (b) The same outcome map can be used for area-selective ALD (ASD)-relevant screening by comparing predicted adsorption and deposition outcomes across candidate growth and non-growth surfaces. Selectivity may arise from inherent surface chemistry differences, where deposition occurs preferentially on the growth surface, or from inhibitor-mediated behavior, where adsorption on the non-growth surface suppresses subsequent film formation.

In this paper, we present AJIN (ALD Joint Interpretable Network), a machine learning framework that predicts ALD adsorption and deposition outcomes within a single, consistent modeling pipeline. We first formulate adsorption, the precursor-substrate chemisorption step, as a binary classification task using three-dimensional structural embeddings of the precursor and substrate together with process temperature. Following the sequential chemistry of ALD, we extend the same embedding-based framework to deposition by additionally incorporating the structural embedding of the reactant, thereby predicting deposition/non-deposition outcomes for a given precursor–substrate–reactant–temperature combination, as shown in Figure 1(a). The key methodological contribution is the use of graph neural networks (GNNs), specifically the Crystal Hamiltonian Graph Neural Network (CHGNet) and the Atomistic Line Graph Neural Network (ALIGNN), to encode the three-dimensional structures of the precursor, substrate, and reactant into fixed-length embedding vectors. All input structures, including molecular precursors, reactant gases, and crystalline substrates, were first geometry-optimized using the PreFerred Potential (PFP), a universal machine learning interatomic potential (MLIP) implemented in the Matlantis platform, ensuring physically relaxed and consistent structural representations across all material types.

The resulting embeddings were concatenated with the scalar process temperature and used as input features for three machine learning classifiers: random forest (RF), XGBoost (XGB), and multi-layer perceptron (MLP). These models were trained on a curated experimental dataset of 560 ALD/area-selective ALD records spanning 76 precursors, 27 substrates, and 20 reactants. By benchmarking CHGNet and ALIGNN embeddings across all three classifiers against classical one-dimensional (1D) descriptor baselines, we systematically evaluate the effects of structural representation and classifier choice on surface-reaction prediction. Beyond predictive performance, AJIN was further evaluated using SHapley Additive exPlanations (SHAP), component ablation, and leave-one-precursor-out (LOPO) cross-validation to assess feature contributions and generalization to unseen precursor chemistries. These analyses also reveal limitations of the present framework, particularly the weak contribution of the scalar temperature representation and the pronounced class imbalance in the adsorption dataset, providing guidance for future ALD dataset construction and model development. This predictive framework also provides a practical basis for ASD-relevant screening, as illustrated in Figure 1(b). By applying the adsorption and deposition models to the same precursor–reactant–temperature condition across multiple candidate substrates, AJIN can generate a substrate-resolved reactivity map of adsorption/non-adsorption and deposition/non-deposition outcomes. Such a map can help identify surface pairs for which growth is predicted on one substrate but suppressed on another, corresponding to ASD-relevant selectivity trends. Importantly, AJIN does not directly predict pattern-level ASD selectivity or explicitly classify inherent versus inhibitor-mediated ASD mechanisms; rather, it provides reaction-level screening information that can be used to prioritize candidate surface combinations for further experimental validation[15] [16] [17] [18]. In addition, laboratory-derived experimental cases were used as a practical comparison to examine model behavior under training-like, substrate-transfer, and broader precursor–substrate transfer conditions.

# 2. Methods

## 2.1 Workflow

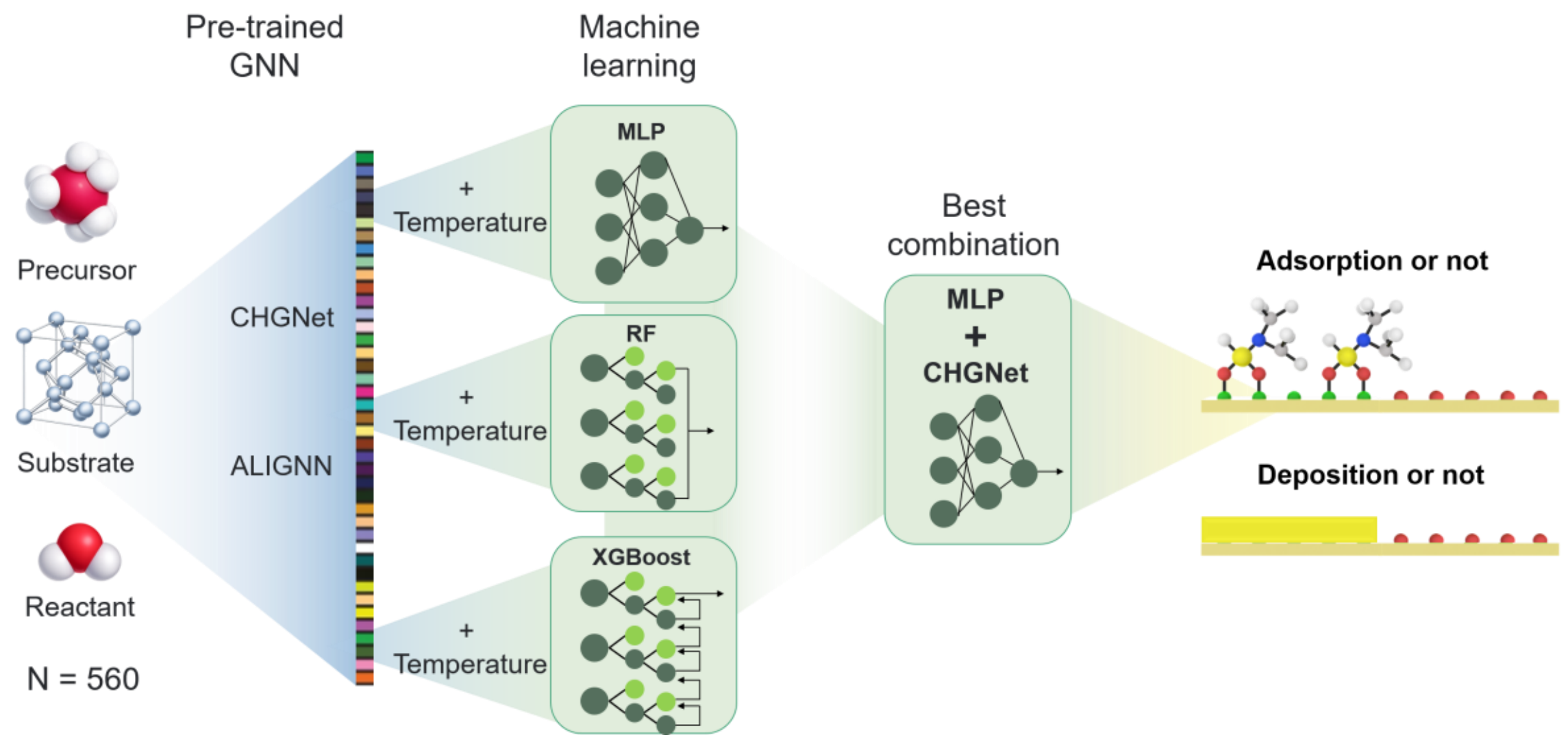

Figure 2. Overview of the AJIN framework for predicting ALD adsorption and deposition outcomes. Precursor, substrate, and reactant structures from a curated dataset of N = 560 experimental ALD/ASD records were geometry-optimized and encoded into fixed-length embedding vectors using two pretrained graph neural networks,

CHGNet and ALIGNN. The resulting embeddings, concatenated with the scalar process temperature, were used to train three machine learning classifiers: MLP, RF, and XGB. Across all embedding-classifier combinations, the CHGNet-MLP configuration achieved the best overall performance and was selected as the champion model. The framework predicts binary adsorption outcomes using precursor, substrate, and temperature inputs, and binary deposition outcomes using precursor, substrate, reactant, and temperature inputs.

The AJIN framework proceeds through four sequential stages, as illustrated in Figure 2: data curation, structure preparation, GNN embedding, and predictive modeling with interpretation. In the first stage, a dataset of 560 experimental ALD/ASD records was curated from published literature. Each record encodes a unique combination of precursor, substrate, reactant, and process temperature, together with separately assigned binary outcome labels for adsorption and deposition (Section 2.2).

In the second stage, all chemical components were prepared as crystallographic information files (CIFs) and geometry-optimized. Substrate representations were derived from Materials Project bulk structures[19], while precursors and reactants were treated as isolated molecules, yielding physically consistent input conformations across all components (Section 2.3).

In the third stage, the relaxed structures were encoded into fixed-length embedding vectors using two pretrained graph neural networks, CHGNet and ALIGNN, which differ in the level of geometric information they capture. A combined embedding strategy, generated by concatenating CHGNet and ALIGNN embeddings, was additionally evaluated. For each record, precursor, substrate, and reactant embeddings were concatenated with the scalar process temperature to form the final input vector (Section 2.4).

In the fourth stage, the assembled vectors were used to train three machine learning classifiers, namely RF, XGB, and MLP. Consistent with the sequential structure of ALD, the framework maintains task-specific input definitions: the adsorption classifier uses precursor, substrate, and temperature features, while the deposition classifier additionally incorporates the reactant embedding. All embedding-classifier combinations were benchmarked under five-fold cross-validation (Sections 2.5 and 2.6), and the CHGNet-MLP configuration achieved the best overall performance and was selected as the champion model. The champion model was further subjected to SHAP analysis, component ablation, and leave-one-precursor-out cross-validation to interpret each prediction task and assess generalization to unseen precursor chemistries (Section 2.7). The two prediction tasks follow the sequential chemistry of the ALD cycle, although the deposition classifier is trained independently using an expanded feature set rather than being explicitly conditioned on the predicted adsorption outcome.

## 2.2 Dataset Construction

| Statistic | Value |
|---|---|
| Total number of samples collected | **560** |
| The number of unique precursors | **76** |
| The number of unique substrates | **27** |
| The number of unique reactants | **20** |
| Temperature range | **25 – 620 °C** |
| Adsorption / non-adsorption cases | **437 / 123 (78.0%)** |
| Deposition / non-deposition cases | **292 / 268 (52.1%)** |

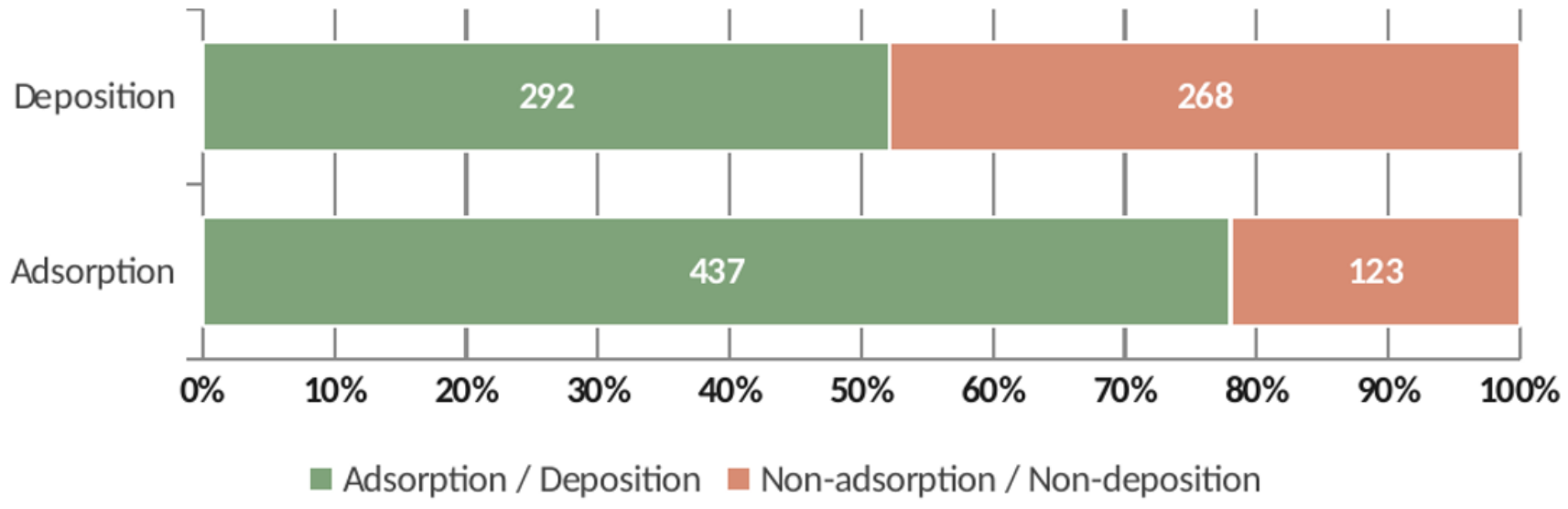

Table 1. Summary statistics of the compiled ALD/ASD experimental dataset (N = 560).

A total of 560 ALD/ASD experimental records were compiled from previously published literature, comprising binary labels for adsorption/non-adsorption and deposition/non-deposition outcomes under diverse process conditions. Each record consists of a unique combination of precursor, substrate, reactant, and process temperature, along with the experimentally assigned binary labels for adsorption/non-adsorption and deposition/non-deposition (1 = adsorption / deposition, 0 = non-adsorption / non-deposition). Adsorption and deposition labels were assigned based on the experimentally reported outcomes of each ALD/ASD process. Adsorption was defined as experimental evidence that the precursor could interact with or chemisorb on the target surface during the precursor exposure, as inferred from reported surface reactions, nucleation behavior, or subsequent process outcomes. Because direct adsorption measurements were not available for every literature record, the adsorption label should be interpreted as an experimentally supported precursor–surface interaction outcome rather than as direct observation of a single elementary chemisorption step. Non-adsorption was assigned when the literature reported no surface adsorption or negligible adsorption based on experimental evidence. When computational evidence, such as density functional theory (DFT), was used to support the assignment, it was considered only in connection with experimentally reported ALD/ASD behavior. Deposition was defined as experimentally confirmed film formation after the complete precursor/reactant cycle, including measurable growth, thickness increase, or reported ALD-type deposition behavior. Non-deposition was assigned when no film growth, negligible growth, or intentionally inhibited deposition was reported under the corresponding process condition. Cases in which the literature reported an initial nucleation delay, that is, a number of early cycles with little or no measurable growth, followed by eventual film formation once steady-state growth was established, were labeled as adsorption and deposition; the transient delay itself was not encoded as a separate outcome, since the final, steady-state result of the process was the basis for labeling. When the original literature did not provide sufficient evidence to distinguish adsorption and deposition outcomes, the record was excluded from the labeled dataset. The dataset encompasses 76 unique precursors, 27 unique substrates, and 20 unique reactants, with process temperatures ranging from 25 to 620 °C, as shown in Table 1. The class distribution exhibits pronounced imbalance in the adsorption task, with 437 adsorption cases (78.0 %) and 123 non-adsorption cases (22.0 %), while the deposition task shows a more balanced distribution with 292 deposition cases (52.1 %) and 268 non-deposition cases (47.9 %). A summary of the dataset statistics is provided in Table 1, and detailed distributions of precursor, substrate, and reactant compositions are provided in Figure S1. The literature sources from which experimental data were collected are listed in the Supporting Information.

## 2.3 Structure Preparation

All input material structures were prepared as independent CIF files prior to embedding. Each component, namely precursor, substrate, and reactant, was prepared as a separate CIF file and processed independently throughout the embedding pipeline. Substrate materials were represented using bulk crystal structures obtained from the Materials Project database, with a 15 Å vacuum region introduced along one cell direction to generate a consistent surface-like structural representation for embedding purposes[8], as shown in Figure S2. No explicit surface cleavage, crystallographic facet selection, or termination-specific slab construction was performed. Precursor and reactant species, which are molecular compounds, were constructed as isolated molecular structures in individual CIF files. All structures, regardless of material type, were subsequently subjected to geometry optimization using PFP, a universal MLIP developed by Matlantis Corporation and accessible via its cloud-based Matlantis platform[20 21], until the maximum force on any atom fell below 0.05 eV/Å[21]. This relaxation step provides locally relaxed and structurally consistent input configurations across the diverse set of precursors, substrates, and reactants considered in this study. The relaxed CIF structures were then used as inputs to the GNN embedding models described in Section 2.4. This vacuum-separated bulk-derived representation was used to generate a consistent surface-like structural fingerprint for embedding generation. It does not explicitly encode crystallographic surface orientation, surface termination, hydroxylation, defect density, amorphous disorder, or other process-dependent surface states, which are treated as limitations of the current framework.

## 2.4 GNN Embedding

Three GNN-based embedding strategies were evaluated to encode the three-dimensional structural information of each input component into fixed-length numerical vectors[22 23]. CHGNet and ALIGNN were selected as the two base embedding models based on three criteria: versatility, ease of use, and structural expressiveness. Both models are pretrained on large-scale crystal structure databases and are applicable to a wide range of material compositions without requiring system-specific fine-tuning, making them well-suited for the chemically diverse precursor-substrate-reactant combinations considered in this study[24 22] . In addition, both models are openly accessible and straightforwardly applicable to arbitrary CIF-format structures, enabling systematic and reproducible embedding generation across all input materials. Although CHGNet and ALIGNN were originally developed for periodic atomistic structures, they can generate fixed-length embeddings from atomistic graphs represented in CIF format. In this study, molecular precursors and reactants were embedded as isolated structures in sufficiently large simulation cells. Therefore, these embeddings are used as structural fingerprints rather than as direct thermodynamic predictions for isolated molecules. Structurally, the two models differ in the level of geometric information they encode: CHGNet represents crystal structures as graphs with atoms as nodes and interatomic bonds as edges[24], encoding structural and electronic information into a 64-dimensional embedding vector, while ALIGNN extends this representation by additionally incorporating bond-angle information through a line graph construction, yielding a 128-dimensional embedding vector[22]. This architectural difference allows a direct comparison of whether higher-order geometric information improves predictive performance for ALD outcome classification[23]. A third strategy, hereafter referred to as the combined embedding, was additionally evaluated by concatenation of the CHGNet and ALIGNN embeddings, resulting in a 192-dimensional vector, to assess whether combining complementary structural representations provides further performance gains. For each experimental record, embeddings were independently generated for the precursor, substrate, and reactant structures using each GNN model. Because adsorption is evaluated before reactant exposure, the adsorption input vector consisted of precursor and substrate embeddings concatenated with the scalar process temperature. This resulted in 129 dimensions for CHGNet-based models (64 × 2 + 1, where the final +1 corresponds to the scalar process temperature), 257 dimensions for ALIGNN-based models (128 × 2 + 1), and 385 dimensions for combined-embedding models (192 × 2 + 1). In contrast, the deposition input vector included precursor, substrate, and reactant embeddings together with the scalar process temperature, resulting in 193 dimensions for CHGNet-based models (64 × 3 + 1), 385 dimensions for ALIGNN-based models (128 × 3 + 1), and 577 dimensions for combined-embedding models (192 × 3 + 1) as summarized in Table 2. The larger input dimensionality of the deposition task than the adsorption task reflects the additional reactant embedding, which encodes the structural information of the oxidant or co-reactant that directly participates in the ligand exchange reaction driving film formation.

| Embedding | Component dimension | Adsorption input | Total | Deposition input | Total |
|---|---|---|---|---|---|
| CHGNet (64 embedding) | 64 | Precursor (64) + Substrate (64) + T (1) | 129 | Precursor (64) + Substrate (64) + Reactant (64) + T (1) | 193 |
| ALIGNN (128 embedding) | 128 | Precursor (128) + Substrate (128) + T (1) | 257 | Precursor (128) + Substrate (128) + Reactant (128) + T (1) | 385 |
| Combined (192 embedding) | 192 | Precursor (192) + Substrate (192) + T (1) | 385 | Precursor (192) + Substrate (192) + Reactant (192) + T (1) | 577 |

**Precursor : precursor embedding**
**Substrate : substrate embedding**
**Reactant : reactant embedding**
**T : temperature**

Table 2. Construction and dimensionality of task-specific input vectors for adsorption and deposition prediction.

## 2.5 Machine Learning Classifiers

Three machine learning classification algorithms were employed to evaluate the predictive utility of the GNN embeddings described in Section 2.4. RF is an ensemble method that constructs many decision trees during training and outputs the majority vote as the final prediction, offering strong resistance to overfitting[25]. XGB is a gradient boosting algorithm that sequentially builds decision trees, with each tree correcting the errors of its predecessor, and is widely recognized for its effectiveness on structured tabular data[26]. MLP is a feedforward neural network consisting of an input layer, one or more hidden layers with non-linear activation functions, and an output layer, capable of learning complex non-linear decision boundaries[27]. Although MLP is a type of neural network, the architecture used in this study consists of only two hidden layers (128 and 64 neurons, respectively) and is therefore categorized as a conventional machine learning classifier rather than a deep learning model, consistent with RF and XGB. All three classifiers were trained and evaluated independently for the adsorption and deposition tasks. Hyperparameter configurations for all three classifiers are summarized in Table S1.

## 2.6 Evaluation Metrics

Model performance was evaluated using multiple complementary metrics. Balanced accuracy (BA), defined as the arithmetic mean of sensitivity and specificity, was adopted as the primary metric to provide an unbiased performance estimate under class-imbalanced conditions[28]:

$$\mathrm{BA} = \frac{\mathrm{Sensitivity} + \mathrm{Specificity}}{2} = \left(\frac{\mathrm{TP}}{\mathrm{TP}+\mathrm{FN}} + \frac{\mathrm{TN}}{\mathrm{TN}+\mathrm{FP}}\right) / 2$$

where TP, TN, FP, and FN denote true positives, true negatives, false positives, and false negatives, respectively. A BA of 50 % corresponds to random chance, while 100 % represents perfect classification. Average precision (AP) was used as a secondary metric, computed as the area under the precision-recall (PR) curve[29]:

$$\mathrm{AP} = \Sigma\,(R_n - R_{n-1})P_n$$

where $P_n$ and $R_n$ denote precision and recall at the n-th threshold, respectively. AP provides a stringent measure of predictive performance particularly suited to imbalanced datasets. The area under the receiver operating characteristic (ROC) curve (AUC-ROC) was additionally reported as a threshold-independent measure of discriminative ability, where 0.5 indicates random classification and 1.0 indicates perfect discrimination. To provide a comprehensive characterization of prediction quality, the following secondary metrics were also computed from the confusion matrix[30]:

$$\text{Sensitivity, true positive rate (TPR)} = \frac{\mathrm{TP}}{\mathrm{TP} + \mathrm{FN}}$$

$$\text{Specificity, true negative rate (TNR)} = \frac{\mathrm{TN}}{\mathrm{TN} + \mathrm{FP}}$$

$$\text{Positive predictive value (PPV)} = \frac{\mathrm{TP}}{\mathrm{TP} + \mathrm{FP}}$$

$$\text{Negative predictive value (NPV)} = \frac{\mathrm{TN}}{\mathrm{TN} + \mathrm{FN}}$$

$$\text{False positive rate (FPR)} = \frac{\text{FP}}{\text{TN} + \text{FP}}$$

These metrics provide a more granular characterization of prediction quality[30 31], particularly relevant under class-imbalanced conditions where overall accuracy can be misleading.

For five-fold cross-validation, the dataset was partitioned at the precursor level such that all experimental records associated with a given precursor were assigned exclusively to a single fold. This precursor-level separation prevented records of the same precursor from appearing in both the training and test sets. The same cross-validation strategy was applied consistently across all embedding–classifier combinations, and performance was reported as the mean and standard deviation across the five held-out folds.

## 2.7 Interpretability and Generalization Analysis

To assess the interpretability of the champion model, SHAP analysis was performed to quantify the contribution of each input feature to individual predictions[32]. SHAP values are derived from cooperative game theory and provide both local (per-sample) and global (dataset-level) explanations of model behavior. Global feature importance was assessed at two levels: at the individual feature level, using the mean absolute SHAP value averaged across all samples to identify the top contributing embedding dimensions; and at the component group level, where SHAP values were aggregated across all dimensions belonging to each task-specific input component. For the adsorption task, the component groups were precursor, substrate, and temperature, whereas for the deposition task, the component groups were precursor, substrate, reactant, and temperature.

To further quantify the contribution of each input component, ablation studies were conducted by systematically removing input components from the task-specific feature vectors[33]. For the adsorption task, precursor, substrate, and temperature components were individually removed. For the deposition task, precursor, substrate, reactant, and temperature components were individually removed. When a component was removed, its corresponding embedding dimensions were entirely excluded from the input feature vector, and the model was retrained from scratch using only the remaining components, while all other settings were kept identical to the full model. The resulting drop in BA and AP relative to the full model was measured to assess the importance of each component. Single-component ablation was performed for both the adsorption and deposition tasks to assess the independent contribution of each input component.

To evaluate generalization to unseen precursor chemistries, leave-one-precursor-out (LOPO) cross-validation was performed across all 76 precursors[31]. This test was designed to answer a question of direct practical relevance to ALD process development: can the model predict outcomes for a precursor it has never seen before, the exact situation a researcher faces when evaluating a newly synthesized or commercially available precursor with no prior ALD literature? In each of the 76 iterations, every experimental record belonging to one precursor was completely removed from the training data and used only as the test set, while the model was retrained from scratch on the remaining 75 precursors. While the 5-fold cross-validation described in Section 2.6 was also split at the precursor level to prevent any single precursor's records from appearing in both training and test folds, each fold there still removes a randomly grouped subset of multiple precursors at once and reports only an aggregate score across that mixed group. LOPO instead isolates exactly one precursor per iteration across all 76 iterations, providing an individual generalization score for every single precursor and thereby allowing a precursor-by-precursor assessment of which chemistries the model transfers to well and which it does not. The same hyperparameter configuration as the champion model identified in the 5-fold cross-validation (Table S1) was used in every iteration, so that the results reflect the model's generalization ability rather than additional tuning. The LOPO procedure is illustrated in Figure S3. BA was summarized across all 76 LOPO iterations for descriptive completeness. Because AP does not provide a meaningful measure for single-class held-out sets, AP was summarized only for precursors containing both outcome classes (42 for adsorption and 35 for deposition). BA for these two-class subsets was also reported to provide a more conservative assessment of precursor-level generalization. For 34 of 76 precursors in the adsorption task and 41 of 76 precursors in the deposition task, all held-out records belonged to a single class. For these folds, BA and AP do not provide a meaningful measure of two-class discrimination and may yield trivial extreme values.

### 2.8 Baseline Models

To assess the value of three-dimensional structural embeddings, two baseline models were constructed for comparison with the champion model. The first was a majority class classifier, which unconditionally predicts the majority label for each task, namely adsorption for the adsorption task and deposition for the deposition task, regardless of input; this establishes a trivial lower bound that any meaningful model must surpass. The second was a classical one-dimensional (1D) descriptor-based model, constructed using molecular descriptors computed with RDKit,34 comprising morphological and geometric properties, physicochemical properties, coordinate statistics, radial distribution function features, elemental fractions, and structural one-hot encodings, combined with process temperature as input features and MLP as the classifier. Following the task-specific input structure used in AJIN, the adsorption baseline used RDKit descriptors of the precursor and substrate together with temperature, whereas the deposition baseline additionally incorporated RDKit descriptors of the reactant. The full list of RDKit-based descriptors is provided in Table S2.

## 3. Results and Discussion

### 3.1 Comparison with Baseline Models

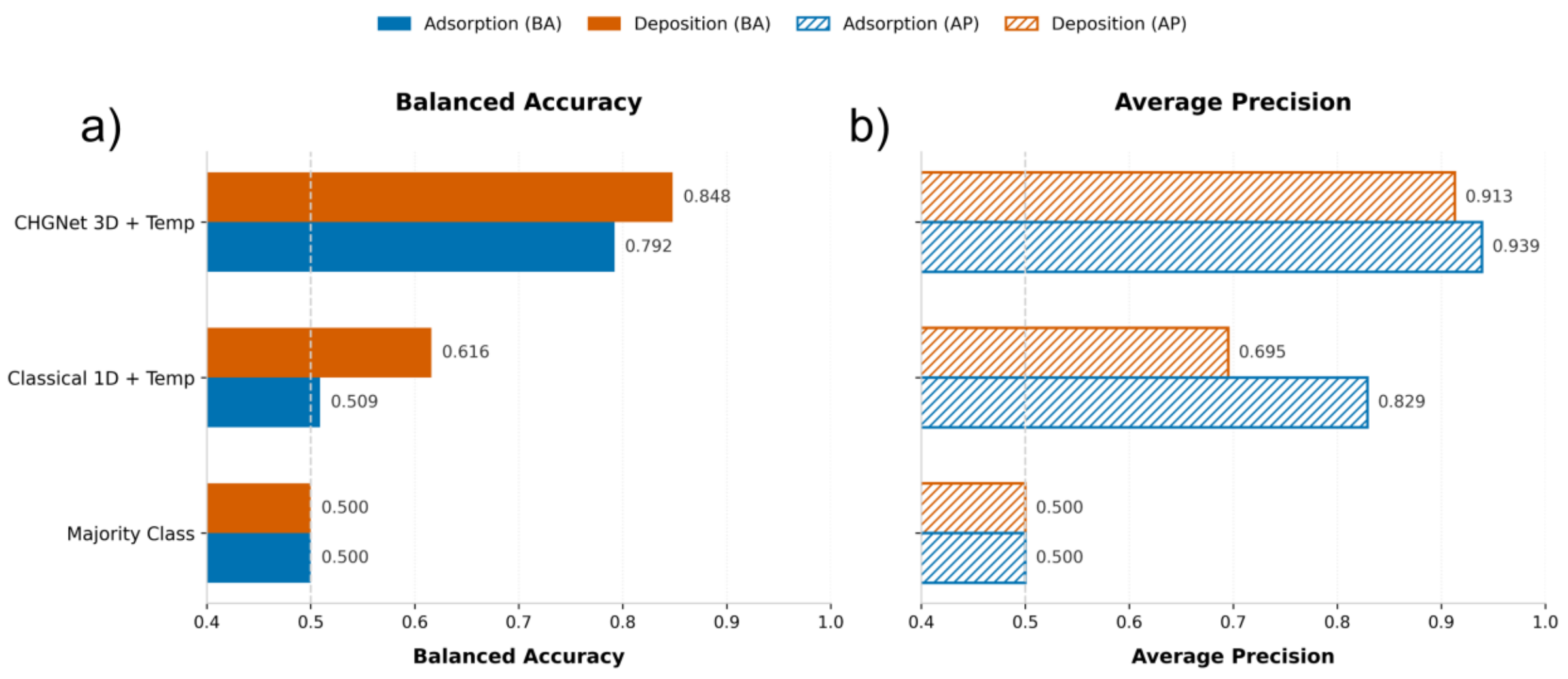

Figure 3. Comparison of (a) BA and (b) AP across three models for adsorption and deposition tasks: a majority class classifier that unconditionally predicts adsorption for the adsorption task and deposition for the deposition task, a classical 1D descriptor-based model using RDKit molecular descriptors with MLP as the classifier, and the champion model using CHGNet 3D structural embeddings with MLP as the classifier. The dashed horizontal line in panel (a) indicates the 50% chance level for balanced accuracy.

To validate the value of three-dimensional structural embeddings, CHGNet-MLP was used as a representative 3D embedding-based model and compared against the majority class classifier and the classical one-dimensional (1D) descriptor-based model described in Section 2.8. CHGNet-MLP was selected for this comparison because it achieved the best overall performance among all GNN embedding-classifier combinations, as detailed in Section 3.2. The majority classifier yields a BA of 50.0 % for both tasks, establishing the trivial lower bound that any meaningful model must surpass. The results are summarized in Figure 3.

For the adsorption task, the classical 1D descriptor-based model achieved a BA of only 50.9 %, marginally above random chance, indicating that the selected RDKit molecular descriptors provide limited discriminative information for predicting adsorption/non-adsorption outcomes in this dataset. In contrast, the champion CHGNet-MLP model achieved a BA of 79.2 %, representing an improvement of 28.3 % points over the classical RDKit baseline. This substantial gap highlights that three-dimensional structural information encoded by CHGNet, including physically informed local bonding environments, coordination geometry, and structural features, provides substantial predictive value for capturing structural information associated with precursor–substrate reactivity. Notably, the classical RDKit descriptor-based model achieved an AP of 82.9 % despite its near-random BA, an apparent discrepancy attributable to the pronounced class imbalance in the adsorption task (78 % positive rate), which can artificially inflate AP even for weak classifiers. The champion model further improved AP to 93.9 %, an increase of 11.0 % points over the classical RDKit baseline, confirming its superior discriminative ability beyond what class imbalance alone can explain.

For the deposition task, the classical 1D descriptor-based model achieved a BA of 61.6 %, showing a more meaningful improvement over random chance than in the adsorption task. This difference should not be interpreted as a direct comparison between two independent tasks, but rather in the context of the sequential ALD framework described in Section 3.2. The deposition task extends the precursor–substrate–temperature input structure used for adsorption by additionally incorporating reactant information, which is chemically required to describe the deposition outcome after reactant exposure. Because deposition depends on whether the adsorbed precursor can react with the co-reactant through ligand exchange or other film-forming reactions, even RDKit-based 1D molecular descriptors can partially capture reactant-dependent chemical trends. However, this descriptor-based signal remains limited compared with the three-dimensional structural representation provided by CHGNet. The champion CHGNet-MLP model achieved a BA of 84.8 % and an AP of 91.3 % for deposition, representing improvements of 23.2 % points in BA and 21.8 % points in AP over the classical RDKit baseline.

Taken together, these baseline comparisons demonstrate that three-dimensional structural embeddings provide substantial and consistent predictive gains over RDKit-based one-dimensional descriptor approaches for both ALD outcome prediction tasks. For adsorption, the large improvement emphasizes the importance of incorporating structural information from both the precursor and substrate. For deposition, the improvement shows that reactant-dependent film formation is also better described when the structural information of precursor, substrate, and reactant is represented in three dimensions. These results support the central design of AJIN: adsorption is modeled from precursor, substrate, and temperature information, and deposition is modeled as a chemically motivated extension that additionally incorporates the reactant embedding.

## 3.2 Benchmarking of Embedding Architectures and Machine Learning Classifiers

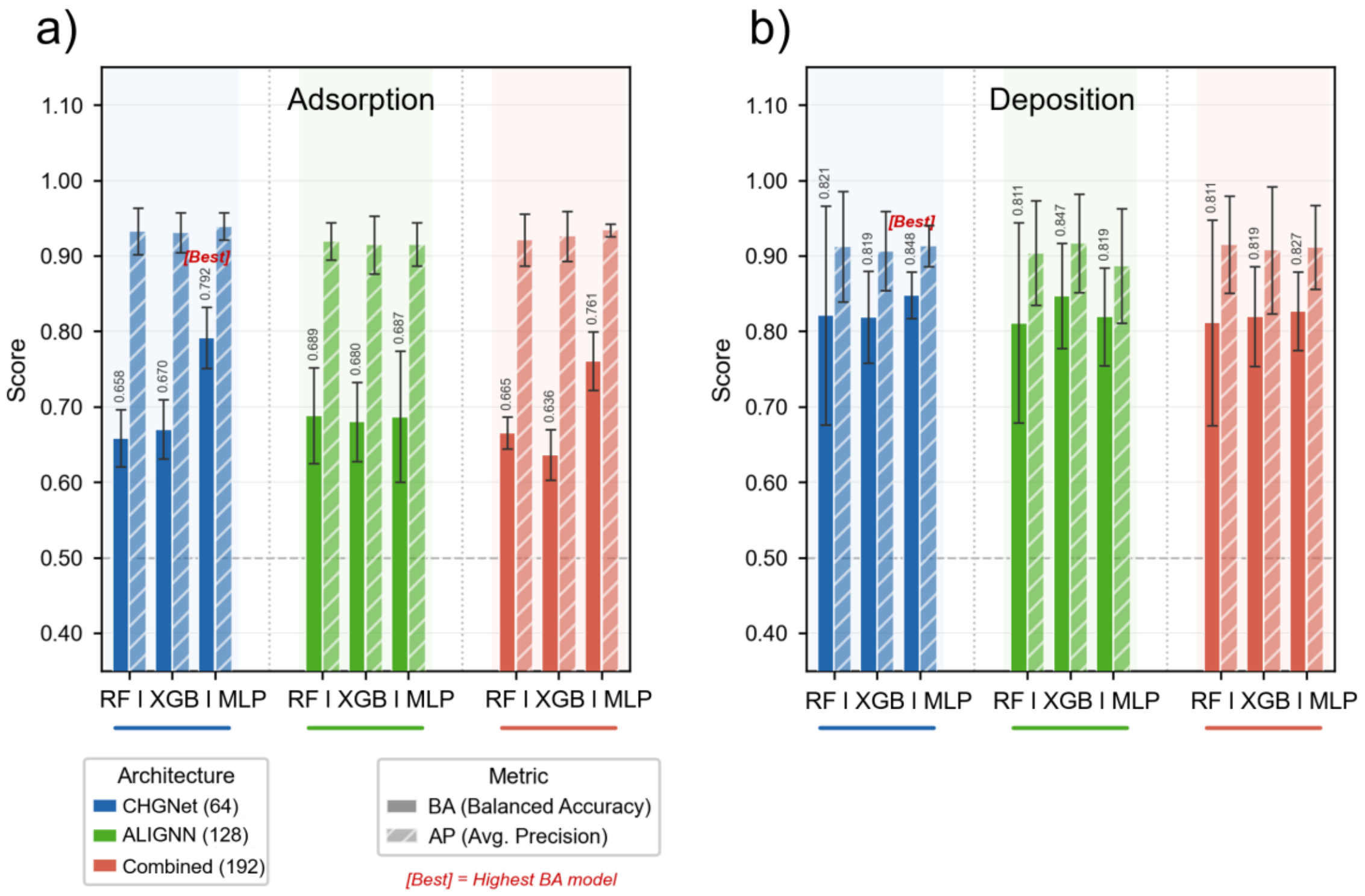

Figure 4. Performance comparison of nine machine learning models across three GNN embedding architectures (CHGNet, ALIGNN, and combined) and three classifiers (RF, XGB, and MLP) for (a) adsorption and (b) deposition binary classification tasks, evaluated by 5-fold cross-validation. Dark and light bars represent BA and AP, respectively. Error bars indicate standard deviation across five folds. [Best] denotes the model with the highest BA for each task.

To identify the optimal combination of structural embedding and classification algorithm, nine models were systematically evaluated by pairing three GNN embedding architectures (CHGNet, ALIGNN, and combined) with three machine learning classifiers (RF, XGB, and MLP) using 5-fold cross-validation. BA was adopted as the primary metric to provide an unbiased performance estimate under class-imbalanced conditions, with AP reported as a complementary measure. All results are reported as mean and standard deviation across five folds. The results are summarized in Figure 4(a) and 4(b) for the adsorption and deposition tasks, respectively, and ROC curves and PR curves for all nine models are provided in Figure S4.

Across all three embedding architectures, MLP consistently outperformed RF and XGB in both BA and AP for both outcome prediction tasks, indicating that the non-linear decision boundaries learned by MLP are better suited to the complex structure–reactivity relationships encoded in GNN embeddings. Among all nine models, CHGNet-MLP achieved the highest performance for both tasks, reaching a BA of 79.2 % and AP of 93.9 % for adsorption, and a BA of 84.8 % and AP of 91.3 % for deposition. Furthermore, the CHGNet-MLP combination exhibited the smallest standard deviation across five folds, indicating superior training stability compared with RF and XGB under the same embedding conditions. These results collectively establish CHGNet-MLP as the champion model for GNN-based ALD outcome prediction.

Regarding embedding architecture, CHGNet consistently achieved the highest BA across all three classifiers for both tasks. This performance gap is likely related to the specific physical information encoded during pretraining. While ALIGNN explicitly incorporates higher-order geometric information through bond-angle representations[22], CHGNet is pretrained to learn energy-, force-, stress-, and magnetic-moment-related information[24]. In the context of ALD, where adsorption and deposition outcomes are strongly influenced by potential energy surfaces, local bonding environments, and surface reactivity, CHGNet embeddings may therefore provide a more physically grounded representation of the input components. The combined embedding, despite integrating CHGNet and ALIGNN vectors into a 192-dimensional representation for each input component, did not surpass CHGNet alone. This result suggests that simply increasing the dimensionality of the input representation does not necessarily improve predictive performance. Instead, the lower performance of the combined embedding is likely attributable to the propagation of less task-relevant or unstable information from the ALIGNN representation and to the increased risk of overfitting associated with the higher-dimensional input relative to the available dataset size of N = 560.

Importantly, the adsorption and deposition tasks should not be interpreted as two independent and directly parallel classification problems. Rather, they reflect the sequentially coupled nature of the ALD cycle. The adsorption task evaluates whether the precursor interacts with the substrate surface before reactant exposure; therefore, its input vector consists of precursor, substrate, and temperature information. In contrast, the deposition task extends the same precursor–substrate–temperature framework by adding the reactant embedding, because film formation is determined after the subsequent reactant exposure. Thus, the deposition model is not an unrelated comparison to the adsorption model, but a chemically motivated extension of the adsorption feature space.

This distinction is important for interpreting the higher performance observed in the deposition task. Across all nine models, the adsorption task achieved a BA range of 63.6–79.2 %, whereas the deposition task achieved a higher BA range of 77.2–84.8 %. This improvement should not be attributed simply to the larger number of input dimensions, because the combined embedding results show that increasing dimensionality alone can degrade performance. Instead, the reactant embedding provides chemically indispensable information for predicting reactant-driven film formation, where the reactant drives ligand exchange and determines whether film formation proceeds after precursor adsorption. This interpretation is consistent with the SHAP analysis presented in Section 3.4, which confirms that reactant embeddings contribute meaningfully to deposition predictions beyond the precursor and substrate contributions already required for adsorption. In addition, the more balanced class distribution in the deposition task (52.1 % positive rate) compared with the adsorption task (78.0 % positive rate) likely allows the model to learn deposition and non-deposition outcome patterns more uniformly, resulting in more robust classification boundaries.

### 3.3 Performance Evaluation of the Champion Model

After CHGNet-MLP was identified as the champion model through the baseline comparison and GNN benchmark analyses, its detailed performance was evaluated using ROC curves, PR curves, and confusion matrices for adsorption and deposition outcome prediction, as presented in Figure 5. These analyses provide complementary threshold-dependent and threshold-independent characterizations of the model's discriminative ability and prediction reliability, complementing the BA and AP metrics reported in Sections 3.1 and 3.2.

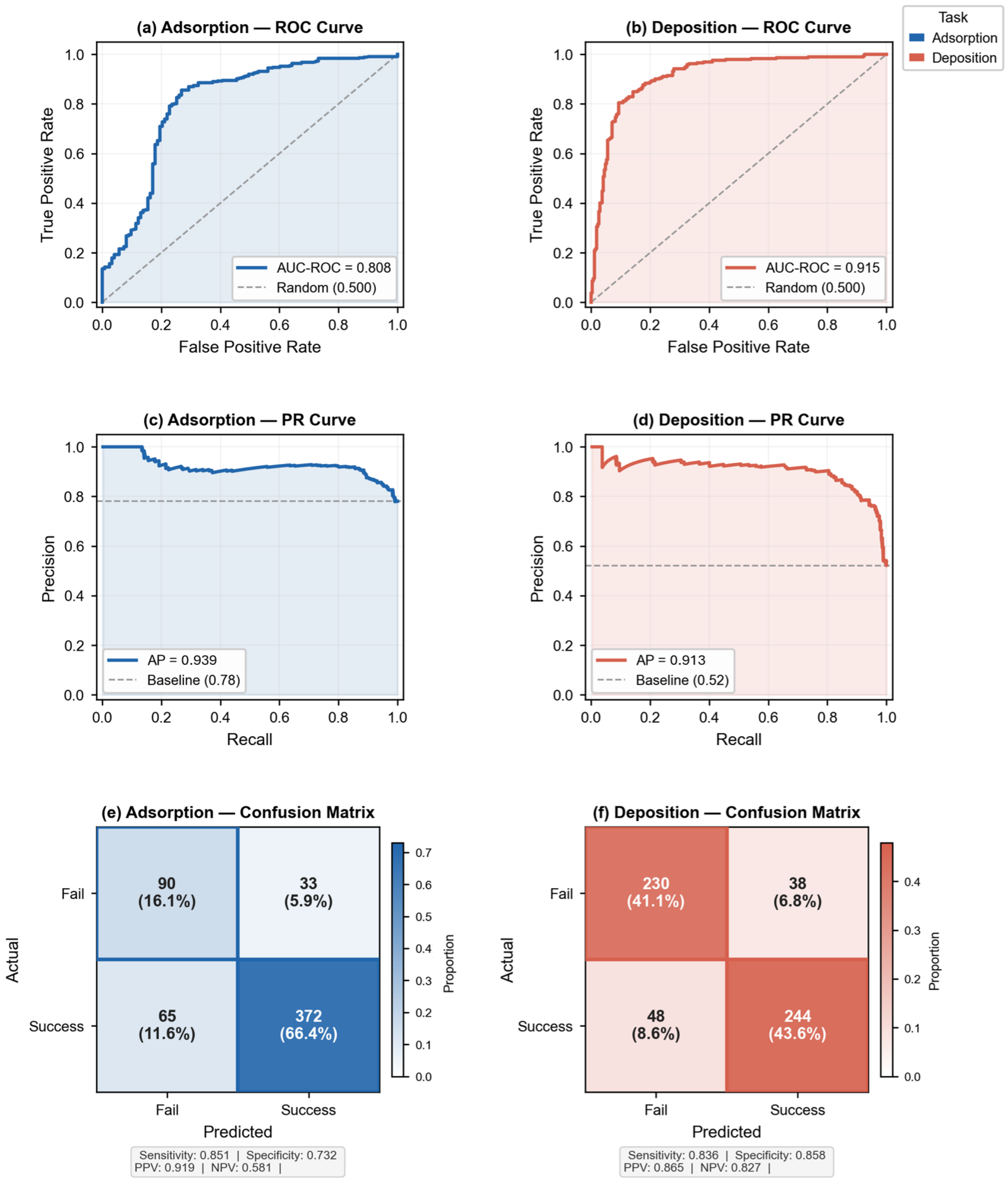

Figure 5. Detailed performance evaluation of the champion model (CHGNet-MLP) for adsorption and deposition tasks. (a, b) ROC curves with AUC-ROC values; the gray dashed line indicates the random classifier baseline (AUC = 0.500). (c, d) Precision-recall curves with AP values; the gray dashed line indicates the adsorption or deposition class prevalence (adsorption: 0.78, deposition: 0.52). (e, f) Confusion matrices showing absolute counts and proportions for adsorption and deposition tasks, respectively. Secondary metrics including sensitivity, specificity, PPV, and NPV are reported below each confusion matrix.

The ROC curves for adsorption and deposition outcome prediction are presented in Figure 5(a) and 5(b), respectively. For adsorption, the ROC curve exhibits a gradual rise in the low-FPR region (FPR = 0–0.2), indicating that a subset of adsorption and non-adsorption cases is difficult to separate under strict classification thresholds. This behavior suggests that some precursor–substrate combinations occupy overlapping regions of the feature space, where the structural signatures associated with surface adsorption are not fully distinct. Beyond FPR = 0.2, the curve rises more rapidly, indicating that additional adsorption cases can be retrieved as the classification threshold is relaxed. The AUC-ROC value of 0.808 indicates that the model captures meaningful discriminative information for adsorption/non-adsorption prediction.

For deposition, the ROC curve rises sharply from the origin, achieving a TPR of approximately 0.8 at FPR below 0.2. Since the deposition input includes precursor, substrate, reactant, and temperature information, this behavior indicates that the model effectively captures the feature patterns associated with deposition and non-deposition outcomes after reactant exposure. The AUC-ROC value of 0.915 indicates strong discriminative ability for deposition/non-deposition prediction.

The precision-recall curves are presented in Figure 5(c) and 5(d). For adsorption, precision remains above 0.9 throughout the recall range of 0–0.8, demonstrating that adsorption predictions made by the model are highly reliable within this operating range. Beyond a recall of 0.8, precision drops sharply, indicating that retrieving the remaining adsorption cases requires accepting more non-adsorption cases as false positives. This pattern reflects the presence of adsorption cases that are difficult to distinguish from non-adsorption cases using the current precursor–substrate–temperature representation.

For deposition, precision remains above 0.9 up to a recall of approximately 0.7 and then declines gradually. This behavior indicates that deposition predictions remain reliable across a broad operating range when precursor, substrate, reactant, and temperature information are considered together. The PR curve therefore supports that the model captures reactant-dependent film formation behavior following precursor adsorption.

The confusion matrices are presented in Figure 5(e) and 5(f). For adsorption, the dominant error mode is false negatives (FN = 65, 11.6 %), where actual adsorption cases are classified as non-adsorption, substantially outnumbering false positives (FP = 33, 5.9 %). This asymmetric error distribution yields a sensitivity of 0.851 and a specificity of 0.732, indicating that the model is more effective at identifying adsorption cases than at identifying non-adsorption cases. The lower specificity and NPV are partly attributable to the pronounced class imbalance in the adsorption dataset (78% positive rate), which makes the negative non-adsorption class less well represented during training. The positive predictive value (PPV = 0.919) and negative predictive value (NPV = 0.581) further reflect this asymmetry, confirming that adsorption predictions are highly reliable, whereas non-adsorption predictions carry greater uncertainty.

For deposition, true positives (TP = 244, 43.6 %) and true negatives (TN = 230, 41.1 %) are distributed in near-equal proportions, with false negatives (FN = 48, 8.6 %) and false positives (FP = 38, 6.8 %) both low and comparable in magnitude. The resulting sensitivity (0.836) and specificity (0.858) indicate that the model identifies both deposition and non-deposition cases with balanced reliability. The PPV (0.865) and NPV (0.827) similarly show that deposition and non-deposition predictions are both reliable. This behavior is consistent with the more even class distribution of the deposition dataset (52 % positive rate) and with the inclusion of reactant information, which is required to describe film formation after reactant exposure.

## 3.4 Model Interpretability: SHAP Analysis

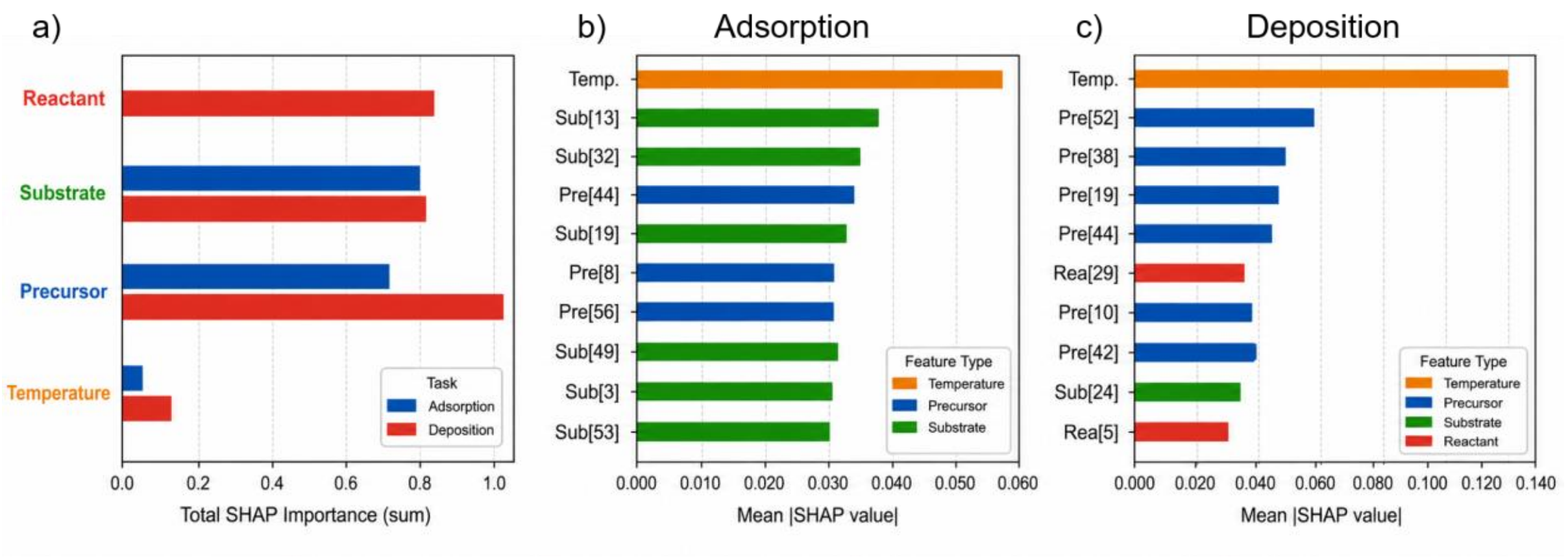

Figure 6. SHAP-based feature importance analysis of the champion model (CHGNet-MLP) for adsorption and deposition tasks. (a) Component group-level SHAP importance, computed as the sum of mean absolute SHAP values across all embedding dimensions within each task-specific input component group. For adsorption, the input groups are precursor, substrate, and temperature; for deposition, the input groups are precursor, substrate, reactant, and temperature. Blue and red bars indicate adsorption and deposition tasks, respectively. (b, c) Individual feature-level SHAP importance showing the top 10 features ranked by mean absolute SHAP value for the adsorption and deposition tasks, respectively. Bar colors indicate feature type: temperature (orange), precursor (blue), substrate (green), and reactant (red).

To examine whether the champion model learns physically meaningful structure–reactivity relationships rather than relying on statistical artifacts, SHAP analysis was performed for the adsorption and deposition prediction tasks. SHAP is a game-theoretic framework that assigns each input feature a contribution value representing its marginal impact on individual predictions, providing both local and global explanations of model behavior. Feature importance was assessed at two complementary levels: the component group level, where SHAP values were aggregated across all embedding dimensions belonging to each task-specific input component, and the individual feature level, where the mean absolute SHAP value of each dimension was computed independently. This two-level analysis is necessary because temperature is represented by a single scalar value, whereas each material component is encoded as a 64-dimensional CHGNet embedding vector. Therefore, group-level and individual-feature-level SHAP values provide different but complementary views of model interpretation. Because the component groups differ substantially in dimensionality, the aggregated SHAP values are interpreted as the total attribution assigned to each component rather than as dimension-normalized feature importance. The ablation analysis in Section 3.5 therefore provides a complementary assessment of the functional contribution of each component to predictive performance. The results are presented in Figure 6, and the SHAP values of the top 40 embedding dimensions, ranked by mean absolute SHAP value, are provided in Figure S5.

At the component group level, the aggregated SHAP importance values show how the model uses different input components for each prediction task, as shown in Figure 6(a). For adsorption prediction, substrate features account for the largest share of total SHAP importance, followed by precursor features, while temperature contributes minimally at the group level. This result is chemically consistent with the surface-mediated nature of precursor adsorption, in which the availability and reactivity of surface sites on the substrate largely determine whether precursor–surface interaction can occur. The precursor contribution further indicates that precursor identity provides additional information related to adsorption propensity once the substrate environment is defined.

For deposition prediction, the model uses precursor, substrate, and reactant information together with temperature. At the component group level, precursor features show the largest contribution, followed by reactant and substrate features with comparable contributions. This ordering is consistent with the chemistry of reactant-driven film formation, where the adsorbed precursor species must undergo ligand exchange or other film-forming reactions with the reactant on the substrate surface. In this context, the precursor embedding captures information related to ligand environment and intrinsic precursor reactivity, the substrate embedding reflects the surface environment established during precursor adsorption, and the reactant embedding provides information directly associated with the subsequent reactant-driven film formation process. Thus, the SHAP analysis supports the task-specific design of AJIN, in which adsorption is modeled using precursor, substrate, and temperature information, while deposition incorporates the reactant as an additional chemically required input.

At the individual feature level, temperature emerges as the single most influential feature for both adsorption and deposition predictions, with mean absolute SHAP values exceeding those of any individual embedding dimension, as shown in Figure 6(b) and 6(c). This result indicates that temperature carries concentrated predictive signal as a scalar variable. However, because each structural component is represented by a distributed 64-dimensional embedding, the importance of a material component is not captured by any single embedding dimension alone. Instead, structural information is distributed across multiple embedding dimensions, which is why the component group-level analysis is necessary for interpreting precursor, substrate, and reactant contributions. This individual-level dominance of temperature, however, contrasts with both the group-level SHAP importance and the ablation results in Section 3.5, where temperature consistently ranks lowest among all input components. Because the ablation results directly measure each component's functional contribution to predictive performance, they should be regarded as the more reliable indicator of temperature's practical importance, consistent with the group-level SHAP ranking.

The remaining top features for both predictions consist primarily of precursor and substrate embedding dimensions, consistent with the chemically important role of precursor–surface interactions in ALD. For deposition prediction, reactant information contributes at both the component group level (Figure 6(a)) and the individual feature level (Figure 6(c)), indicating that the model actively uses reactant structural information when evaluating reactant-driven film formation. This is chemically meaningful because reactant exposure drives ligand exchange and film-forming chemistry after precursor adsorption. However, the apparent dominance of temperature at the individual feature level, together with its small contribution at the component group level, indicates that temperature acts as a concentrated scalar predictor rather than a distributed structural representation. This suggests that the current scalar temperature input may not fully capture the thermodynamic and kinetic roles of temperature across chemically diverse ALD systems. This issue is further examined through the ablation study in Section 3.5.

## 3.5 Ablation Study

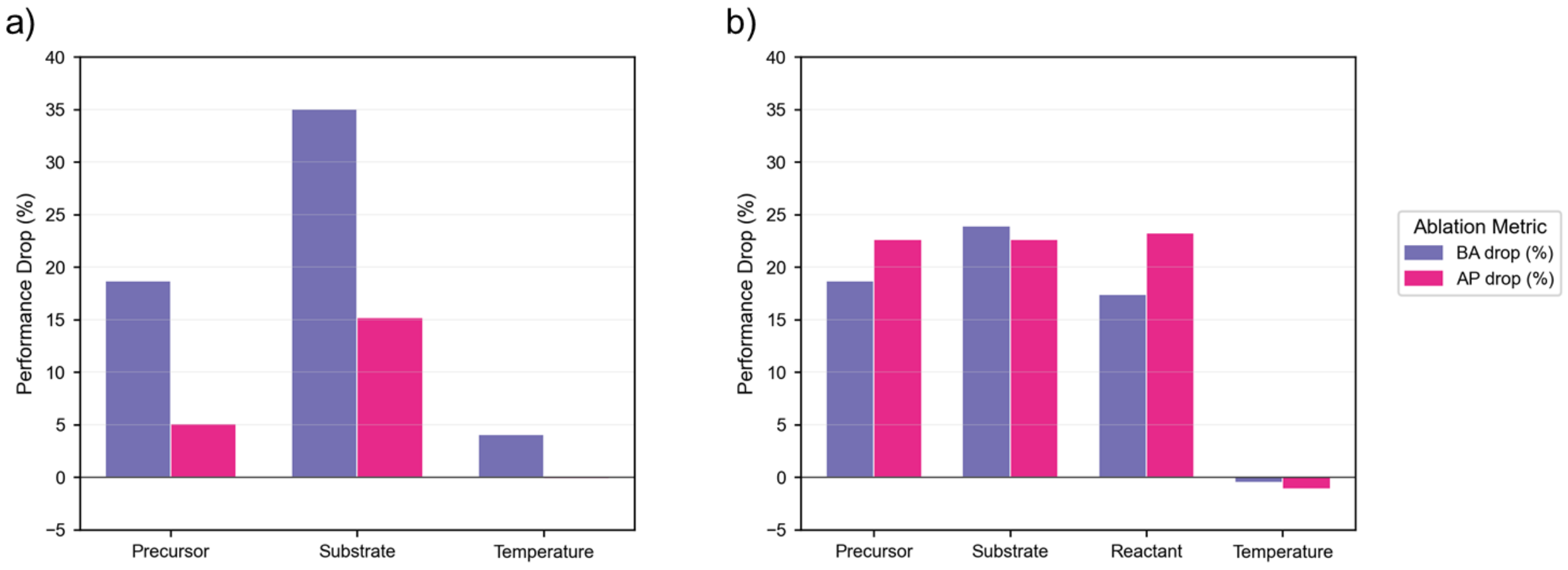

Figure 7. Ablation study results quantifying the independent contribution of each input component to model performance. (a) Single-component ablation for the adsorption task, showing BA and AP drops upon removal of precursor, substrate, and temperature. (b) Single-component ablation for the deposition task, showing BA and AP drops upon removal of precursor, substrate, reactant, and temperature. Positive values indicate performance drops relative to the full model; negative values indicate marginal performance gains upon component removal.

To quantify the functional contribution of each input component and validate the feature importance trends identified by SHAP analysis, ablation studies were conducted by systematically removing each input component and retraining the model from scratch under otherwise identical conditions. For adsorption prediction, precursor, substrate, and temperature components were individually removed from the precursor–substrate–temperature input representation. For deposition prediction, precursor, substrate, reactant, and temperature components were individually removed from the precursor–substrate–reactant–temperature input representation. The resulting changes in BA and AP relative to the full model were used to assess the independent contribution of each component. The results are presented in Figure 7(a) and 7(b).

As shown in Figure 7(a), substrate removal produced the largest performance drop for adsorption prediction, with BA decreasing by 35.0 percentage points and AP by 15.2 percentage points. This reduction brings the adsorption BA to approximately 44%, below the 50% chance level, indicating that the remaining precursor–temperature features alone do not provide reliable discrimination under the present evaluation setting. This result is consistent with the SHAP analysis in Section 3.4, where substrate features showed the largest group-level contribution for adsorption prediction. Chemically, this indicates that adsorption cannot be determined from precursor identity alone; the same precursor can exhibit different adsorption behavior depending on the substrate surface environment. Therefore, substrate surface chemistry, including the availability and reactivity of surface sites, is a primary factor governing precursor adsorption.

Precursor removal also reduced adsorption performance, with BA decreasing by 18.7 percentage points and AP by 5.1 percentage points. This confirms that precursor identity provides additional and independent information for adsorption prediction once the substrate environment is defined. In contrast, temperature removal resulted in only a small BA drop of 4.0 percentage points, while AP remained nearly unchanged. This behavior is consistent with the low group-level SHAP importance of temperature and suggests that the current scalar temperature representation provides limited additional discriminative information for adsorption/non-adsorption prediction in the present dataset.

As shown in Figure 7(b), the deposition ablation results quantify the contribution of each component after reactant information is incorporated for deposition prediction. Substrate removal produced the largest BA drop, with BA decreasing by 23.9 percentage points and AP by 22.7 percentage points. This indicates that the substrate environment remains important for deposition prediction, because film formation after reactant exposure still depends on the surface state established during precursor adsorption. Precursor removal also caused a substantial decrease in performance, with a BA drop of 18.7 percentage points and an AP drop of 22.6 percentage points, indicating that precursor structure and ligand environment provide important information for evaluating whether the adsorbed precursor can participate in film-forming reactions.

Reactant removal resulted in a BA drop of 17.4 percentage points and an AP drop of 23.3 percentage points. This confirms that reactant structural information contributes independently to deposition/non-deposition prediction. Chemically, this is consistent with the role of the reactant in driving ligand exchange and subsequent film formation after precursor adsorption. Therefore, the ablation result supports the task-specific design of AJIN, in which deposition prediction includes reactant information in addition to precursor, substrate, and temperature. Temperature removal yielded a marginal performance gain (BA change: -0.5 percentage points; AP change: -1.1 percentage points), further indicating that the current scalar temperature representation may act as noise rather than informative signal for deposition prediction, given the marginal performance improvement observed upon its removal.

The different responses of BA and AP to component removal also reflect the class distributions of the two prediction tasks. For adsorption prediction, AP drops are smaller than BA drops upon component removal, which is partly attributable to the high adsorption-class prevalence of the adsorption dataset (78 %). In contrast, for deposition prediction, AP drops are comparable to or larger than BA drops, consistent with the more balanced deposition-class prevalence of the deposition dataset (52 %). This result highlights the importance of interpreting BA and AP together when evaluating ALD outcome prediction models trained on datasets with different class distributions.

Taken together, the ablation results provide functional validation of the task-specific input design used in AJIN. For adsorption prediction, substrate and precursor information are both required to describe precursor–surface interaction, with substrate information playing the dominant role. For deposition prediction, precursor, substrate, and reactant information each make non-redundant contributions to predicting deposition/non-deposition outcomes after reactant exposure. In both cases, the limited contribution of the scalar temperature input indicates that temperature is represented only coarsely in the current framework. This limitation motivates future work incorporating denser temperature sampling for fixed precursor–substrate–reactant combinations and more physically meaningful thermodynamic or kinetic descriptors. The generalization ability of the model to entirely unseen precursor chemistries is further examined through leave-one-precursor-out cross-validation in Section 3.6.

## 3.6 Leave-One-Precursor-Out Generalization Analysis

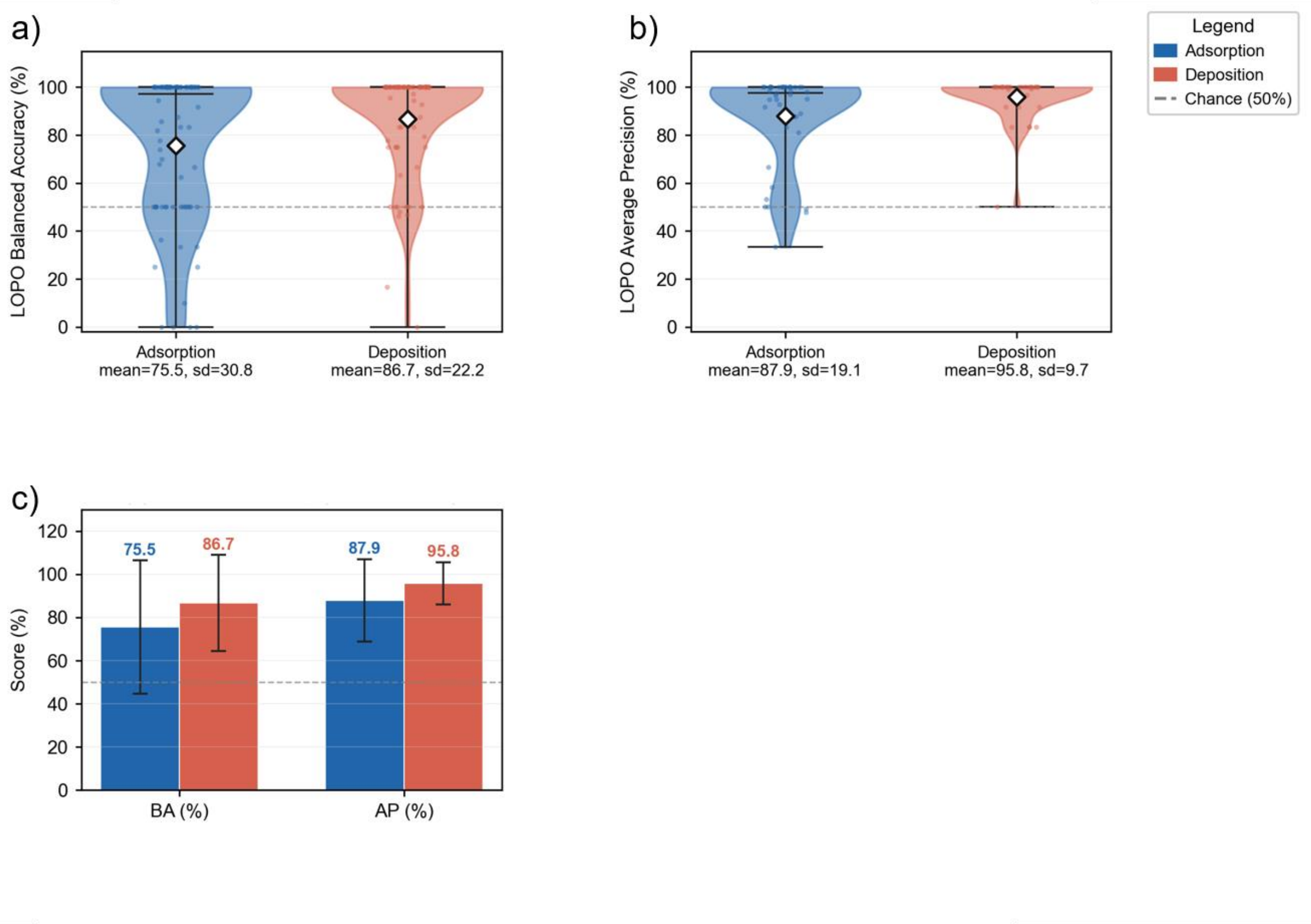

d)

| Case | Precursor | Substrate | Reactant | T (°C) | Real Ads | Pred Ads (%) | Real Dep | Pred Dep (%) | Reference |
|---|---|---|---|---|---|---|---|---|---|
| 1 | TEMAHf | $SiO_2$ | $H_2O$ | 250 | 1 | **99.6** | 1 | **96.7** | 3, 35 |
| 2 | $TiCl_4$ | Si/H-Si | $H_2O$ | 100 | 1 | **98.7** | 0 | **~0** | 18 |
| 3 | $RuO_4$ | $SiO_2$ | $H_2$ | 125 | 0 | **3.3** | 0 | **~0** | 36 |
| 4a | DMADMS | $SiO_2$ | $H_2O$ | 100 | 1 | **100.0** | 0 | **~0** | 37 |
| 4b | DMADMS | $SiO_2$ | $O_3$ | 100 | 1 | **100.0** | 0* | **87.1** | 37 |
| 4c | DMADMS | $SiO_2$ | $H_2O$ | 200 | 1 | **100.0** | 0 | **~0** | 37 |
| 4d | DMADMS | $SiO_2$ | $O_3$ | 200 | 1 | **100.0** | 1 | **99.5** | 37 |
| 4e | DMADMS | $SiO_2$ | $H_2O$ | 300 | 1 | **100.0** | 0 | **~0** | 37 |
| 4f | DMADMS | $SiO_2$ | $O_3$ | 300 | 1 | **100.0** | 1 | **100.0** | 37 |

Correct prediction (Pred > 90%) | Correctly predicted failure | Ambiguous / False positive (4b) | DMADMS condition series

Figure 8. Generalization performance of the champion model (CHGNet-MLP) assessed by leave-one-precursor-out (LOPO) cross-validation across all 76 precursors.(a) BA and (b) AP distributions of per-precursor LOPO results for the adsorption and deposition tasks; diamond markers indicate mean values and the dashed horizontal line denotes the 50% chance level for BA and should not be interpreted as a chance baseline for AP. (c) Summary bar chart showing mean and standard deviation of BA and AP for both tasks. (d) Representative case studies comparing experimentally observed and model-predicted outcomes for adsorption and deposition. Case 1

represents a system classified as adsorption and deposition. Case 2 represents a system classified as adsorption and non-deposition, demonstrating the model's ability to distinguish adsorption/non-adsorption from deposition/non-deposition outcomes. Case 3 represents a system classified as non-adsorption and non-deposition. Cases 4a–4f represent a systematic condition series in which reactant type and process temperature are varied for the same precursor-substrate combination, testing the model's sensitivity to these variables. Cell colors indicate prediction outcome: blue indicates correctly predicted adsorption or deposition outcomes with predicted probability above 90%, pink indicates correctly predicted non-adsorption or non-deposition outcomes, and yellow indicates ambiguous cases warranting experimental re-examination. In panels (a) and (b), extreme values are primarily associated with precursors having single-class held-out records, for which BA and AP do not provide a meaningful measure of two-class discrimination. These folds are therefore retained only for descriptive completeness, while the two-class held-out subsets are used to provide a more conservative assessment of precursor-level generalization (see Section 3.6).

While the preceding precursor-grouped five-fold cross-validation evaluates generalization to held-out precursors, LOPO cross-validation provides a complementary assessment by resolving performance for each precursor individually.. To directly evaluate this out-of-distribution (OOD) generalization capability, leave-one-precursor-out (LOPO) cross-validation was performed across all 76 precursors in the dataset. In each of the 76 iterations, all samples associated with one precursor were withheld as the test set, while the model was retrained from scratch on the remaining data. This protocol ensures complete precursor-level separation between training and test sets and simulates the practically relevant scenario of predicting ALD outcomes for a novel precursor chemistry not represented in prior experimental records. The results are presented in Figure 8.

The LOPO results, summarized in Figure 8(a), 8(b), and 8(c), show that the champion model retains meaningful predictive capability for unseen precursor chemistries. For adsorption prediction, the mean BA across all 76 precursors was 75.5% with a standard deviation of 30.8%. Among the 42 precursors with both outcome classes represented in the held-out set, the mean BA was 66.4% (SD 26.1%) and the mean AP was 87.9% (SD 19.1%). Of the 76 precursors, 34 (44.7%) had held-out records belonging to a single class, for which BA and AP do not provide a meaningful measure of two-class discrimination. These values indicate that the model captures transferable precursor–substrate interaction trends beyond simple precursor-specific memorization. The large standard deviation further shows that adsorption prediction is sensitive to the chemical novelty of the held-out precursor. The violin plot in Figure 8(a) shows that while many precursors yield high LOPO performance, a subset of chemically isolated precursors yields performance near or below the 50% chance level. Overall, generalization performance varied substantially across precursor chemistries, indicating that some held-out chemistries are more difficult to extrapolate than others.

For deposition prediction, the mean BA across all 76 precursors was 86.7% with a standard deviation of 22.2%. Among the 35 precursors with both outcome classes represented in the held-out set, the mean BA was 78.9% (SD 20.7%) and the mean AP was 95.8% (SD 9.7%). Similarly, 41 of 76 precursors (53.9%) had single-class held-out sets, for which two-class discrimination cannot be meaningfully assessed. Since deposition depends on reactant-driven film formation following precursor adsorption, the LOPO result suggests that the model learns broader structure–reactivity relationships involving precursor–substrate–reactant combinations, rather than relying solely on precursor-specific patterns. The remaining precursor-to-precursor variance indicates that generalization remains uneven across precursor chemistries.

To provide concrete illustrations of the model's OOD generalization behavior, representative case studies are presented in Figure 8(d), spanning three distinct precursor chemistries and a systematic condition series based on the DMADMS/$SiO_2$ system. Case 1 (TEMAHf/$SiO_2$/$H_2O$, 250 °C) represents a well-established ALD system within a typical process window[3 35]. The model predicts adsorption and deposition outcomes with high probabilities of 99.6% and 96.7%, respectively, consistent with the experimentally observed ALD behavior. Case 2 ($TiCl_4$/Si-H/$H_2O$, 100 °C) represents a sequentially informative adsorption/non-deposition case[18]. The model predicts a high adsorption probability of 98.7% and a near-zero deposition probability, consistent with inhibited $TiO_2$ nucleation on hydrogen-terminated silicon surfaces. Case 3 ($RuO_4$/$SiO_2$/$H_2$, 125 °C) represents a non-adsorption/non-deposition case[36], and the model predicts low probabilities for both adsorption (3.3%) and deposition, consistent with the reported instability of the $RuO_4$/$H_2$ process on $SiO_2$ at this temperature.

Cases 4a through 4f systematically vary the reactant and temperature for the DMADMS/$SiO_2$ system, providing a controlled test of the model's ability to describe reactant- and temperature-dependent deposition/non-deposition behavior[37]. The model predicts non-deposition for $H_2O$ across the tested temperature range (Cases 4a, 4c, and 4e), whereas $O_3$ leads to deposition at elevated temperatures (Cases 4d and 4f). This behavior is consistent with the role of $O_3$ as a stronger oxidant for enabling the film-forming reaction in this system. Case 4b (DMADMS/$SiO_2$/$O_3$, 100 °C) illustrates a limitation of the current model in capturing transition-sensitive temperature effects. Although temperature appears as a high-impact individual feature in the SHAP analysis, ablation studies showed that removing the scalar temperature input produces only a marginal change in overall model performance. This discrepancy suggests that the current scalar temperature representation may not sufficiently capture sharp, non-linear kinetic thresholds or process-window transitions in certain ALD systems. Therefore, the disagreement observed in Case 4b may arise not only from the ambiguity of the experimental label, but also from the limited ability of the present feature representation to describe temperature-dependent reaction kinetics.

These case studies further suggest that the same prediction framework can be applied to ASD-relevant screening. In ASD, selectivity arises from differences in adsorption/non-adsorption and deposition/non-deposition behavior across growth and non-growth surfaces. By evaluating a given precursor–reactant–temperature condition across multiple substrates or inhibited surfaces, AJIN can identify whether adsorption is expected to occur preferentially on a target surface, whether that adsorption is likely to lead to deposition, and whether adsorption is likely to proceed to deposition or remain as non-deposition. Therefore, the sequential prediction of adsorption/non-adsorption followed by deposition/non-deposition provides a practical route for screening ASD-relevant precursor–substrate–reactant combinations without requiring a separately trained ASD-specific model at this stage.

This result suggests that incorporating more expressive thermodynamic and kinetic descriptors will be important for future iterations of AJIN, potentially enabling a more comprehensive and process-aware mapping of ALD/ASD reaction space. Collectively, the LOPO analysis and representative case studies show that the model captures chemically meaningful and physically interpretable structure–reactivity relationships for unseen precursor chemistries, supporting its practical utility as a screening framework for ALD process discovery and its potential extensibility toward ASD process design.

### 3.7 Comparison with laboratory-derived experimental results

To further examine the practical applicability of AJIN beyond literature-derived benchmark validation, the champion CHGNet-MLP model was applied to laboratory-derived experimental cases grouped according to their relationship to the training dataset (Figure 9). The cases were divided into three transfer priority groups and all experiments were conducted in our group. P0 represents cases identical to the training data, P1 represents cases in which the precursor is retained but the substrate is changed, and P2 represents cases in which both the precursor and substrate differ from the training data. This grouping provides a simple experimental hierarchy for evaluating model behavior from training-like conditions to increasingly challenging transfer conditions. Because the number of cases and chemical diversity differ across these groups, the accuracy values in Figure 9(a) should be interpreted together with the corresponding sample counts shown on each bar.

The P0 group showed 100% accuracy for both adsorption and deposition predictions. However, this result should be viewed primarily as a consistency check rather than evidence of extrapolative capability, because P0 corresponds to conditions already represented in the training distribution. In contrast, P1 provides a more meaningful transfer test because the precursor is retained while the substrate is changed. For this group, adsorption and deposition predictions both reached 83.3% accuracy, corresponding to five correct predictions out of six evaluated cases for each task. This result suggests that AJIN can transfer learned precursor reactivity to related substrate environments to some extent, while also showing that prediction errors begin to emerge when the surface chemistry differs from the original training condition.

The P2 group represents the most chemically diverse and challenging set because both the precursor and substrate differ from the training data. In this group, adsorption accuracy was 71.4%, corresponding to 10 correct predictions out of 14 cases. This lower value, together with the broader diversity of precursor and substrate combinations in P2, indicates that adsorption prediction becomes more uncertain when both molecular and surface chemistry are changed simultaneously. Deposition prediction in P2 showed 100% accuracy, but this value was calculated from only three labeled deposition cases. Therefore, the P2 deposition result should not be interpreted as broad generalization across the full P2 chemical space, but rather as a limited positive indication that the model can capture deposition/non-deposition trends for selected experimentally labeled transfer cases.

The row-level results in Figure 9(b) provide additional context beyond the aggregate percentages. Correct predictions are concentrated in P0 and P1, where the test conditions remain close to the training distribution or preserve the same precursor identity. In contrast, incorrect or ambiguous predictions appear more frequently in P2, where both precursor and substrate chemistry are changed. This pattern is consistent with the leave-one-precursor-out analysis, which showed greater variability for adsorption than for deposition. Overall, the laboratory-derived comparison supports the use of AJIN as a screening tool for prioritizing precursor–substrate–reactant conditions, while also emphasizing that aggregate accuracy must be interpreted in light of sample size, label availability, and chemical diversity. Broader experimental coverage, especially for P2-type transfer cases with both adsorption and deposition labels, will be important for more rigorously evaluating extrapolative performance.

a)

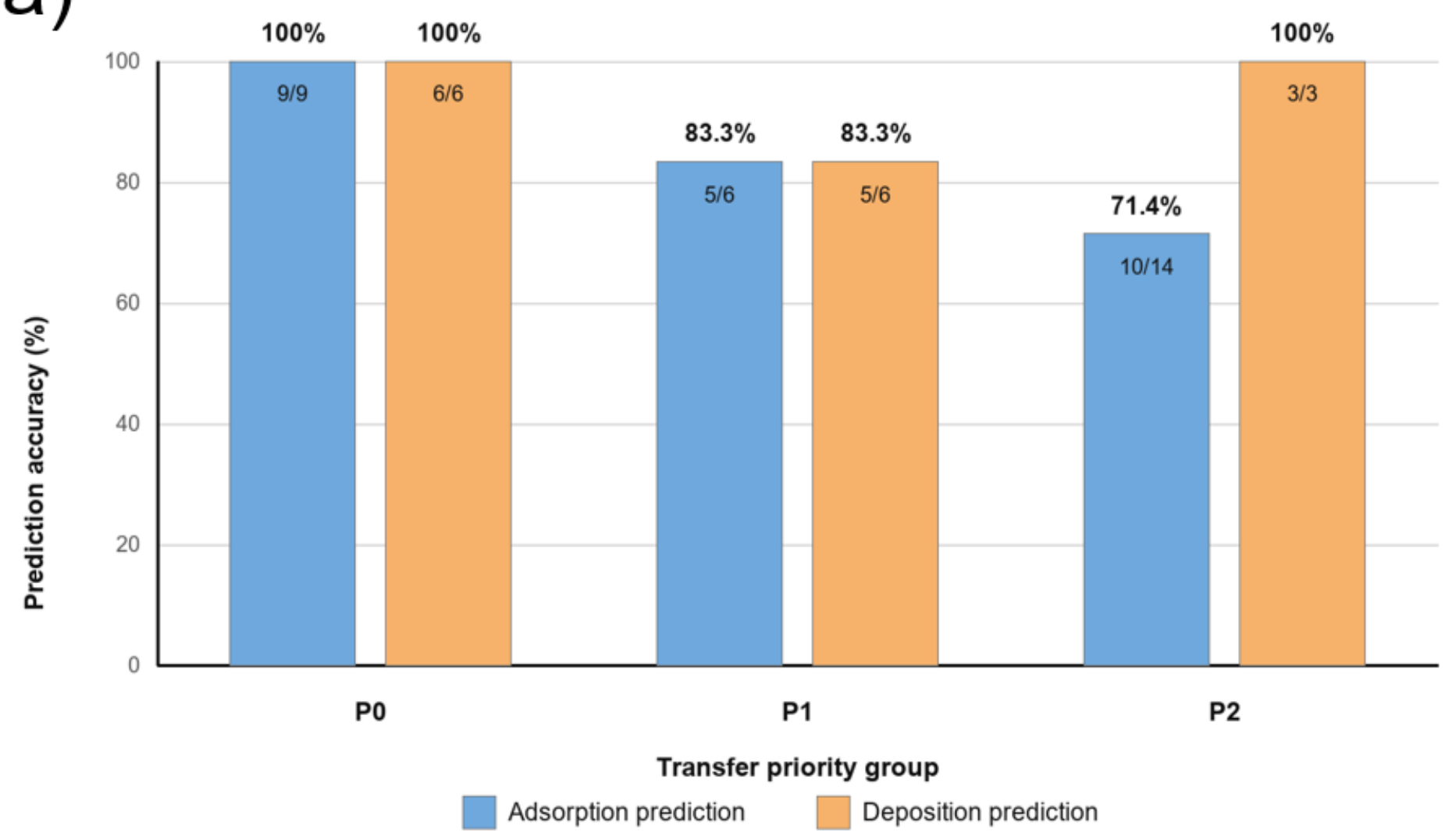

b)

| Case | Precursor | Substrate | Reactant | T (°C) | Real Ads | Pred Ads (%) | Real Dep | Pred Dep (%) | Reference |
|---|---|---|---|---|---|---|---|---|---|
| **P0-1** | bis(N,N-dimethylamino)dimethylsilane | $SiO_2$ | $H_2O$ | 150 | 1 | **98.6** | 0 | **3.8** | 37 |
| **P0-2** | (N,N-dimethylamino)trimethylsilane | $SiO_2$ | $H_2O$ | 150 | 1 | **97.9** | 0 | **5.1** | 37 |
| **P0-3** | (N,N-dimethylamino)trimethylsilane | H-Si | $H_2O$ | 150 | 0 | **7.4** | 0 | **8.2** | 37 |
| **P0-4** | ethanethiol | Co | . | 250 | 1 | **96.8** | . | . | 38 |
| **P0-5** | ethanethiol | Cu | . | 250 | 1 | **99.1** | . | . | 38 |
| **P0-6** | ethanethiol | $SiO_2$ | . | 250 | 0 | **4.6** | . | . | 38 |
| **P0-7** | trimethylaluminum | $SiO_2$ | $H_2O$ | 250 | 1 | **98.2** | 1 | **96.4** | 38 |
| **P0-8** | dimethylaluminum isopropoxide | $SiO_2$ | $H_2O$ | 250 | 1 | **95.7** | 1 | **97.6** | 38 |
| **P0-9** | tetrakis(dimethylamido)titanium(IV) | $SiO_2$ | $H_2O$ | 180 | 1 | **99.4** | 1 | **98.1** | 39 |
| **P1-1** | bis(N,N-dimethylamino)dimethylsilane | H-Si | $H_2O$ | 150 | 0 | **14.8** | 0 | **42.6** | 37 |
| **P1-2** | titanium(IV) isopropoxide | $SiO_2$ | $H_2O$ | 180 | 1 | **88.9** | 1 | **91.7** | 39 |
| **P1-3** | tricarbonyl(trimethylenemethane)ruthenium | $SiO_2$ | $H_2O$ | 200 | 1 | **86.3** | 1 | **89.5** | 40 |
| **P1-4** | tricarbonyl(trimethylenemethane)ruthenium | $Si_3N_4$ | $H_2O$ | 200 | 0 | **63.1** | 0 | **18.4** | 40 |
| **P1-5** | bis(cyclopentadienyl)cobalt(II) | Cu | $NH_3$ plasma | 250 | 1 | **92.8** | 0 | **21.7** | 41 |
| **P1-6** | bis(cyclopentadienyl)cobalt(II) | $SiO_2$ | $NH_3$ plasma | 250 | 1 | **11.2** | 1 | **84.9** | 41 |
| **P2-1** | trimethoxy(pentamethylcyclopentadienyl)titanium(IV) | $TiO_2$ | $H_2O$ | 180 | 1 | **78.6** | 0 | **26.8** | 39 |
| **P2-2** | trimethoxy(pentamethylcyclopentadienyl)titanium(IV) | $SiO_2$ | $H_2O$ | 180 | 1 | **36.9** | 0 | **31.4** | 39 |
| **P2-3** | organosulfide small-molecule inhibitor | Cu | | 400 | 1 | **74.2** | . | . | 42 |
| **P2-4** | organosulfide small-molecule inhibitor | $SiO_2$ | | 400 | 1 | **81.6** | . | . | 42 |
| **P2-5** | organosulfide small-molecule inhibitor | TiN | | 400 | 0 | **19.8** | . | . | 42 |
| **P2-6** | cyclohexane carboxaldehyde | $Si_3N_4$ | | 200 | 1 | **44.7** | . | . | 40 |
| **P2-7** | cyclohexane carboxaldehyde | $SiO_2$ | | 200 | 0 | **61.5** | . | . | 40 |
| **P2-8** | di-n-propyl sulfide | Cu | | 400 | 1 | **83.7** | . | . | 42 |
| **P2-9** | di-n-propyl sulfide | $SiO_2$ | | 400 | 1 | **79.3** | . | . | 42 |
| **P2-10** | di-n-propyl sulfide | TiN | | 400 | 0 | **23.6** | . | . | 42 |
| **P2-11** | diisopropyl sulfide | Cu | | 400 | 1 | **77.8** | . | . | 42 |
| **P2-12** | diisopropyl sulfide | $SiO_2$ | | 400 | 0 | **58.9** | . | . | 42 |
| **P2-13** | diisopropyl sulfide | TiN | | 400 | 0 | **16.5** | . | . | 42 |
| **P2-14** | tetrakis(dimethylamido)titanium(IV) | $TiO_2$ | $H_2O$ | 180 | 1 | **82.4** | 1 | **76.2** | 39 |

Correct positive prediction · Correctly predicted failure · Wrong / ambiguous prediction · P2 transfer-test series

Figure 9. Transfer prediction performance for laboratory experimental cases. (a) Summary of adsorption and deposition prediction accuracy for each transfer priority group. Bar labels indicate the number of correctly predicted cases over the total number of evaluated cases. (b) Row-level prediction table showing the precursor, substrate, reactant, temperature, experimental adsorption/deposition labels, and prediction confidence values. P0 denotes cases identical to the training data, P1 denotes transfer cases with the same precursor but a different substrate, and P2 denotes transfer cases in which both the precursor and substrate differ from the training data. Blue cells indicate correctly predicted positive cases, pink cells indicate correctly predicted negative cases, yellow cells indicate wrong or ambiguous predictions, and gray shading marks the P2 rows used for broader precursor–substrate transfer testing.

## Summary

In this study, we developed AJIN, an interpretable machine learning framework for predicting adsorption/non-adsorption and deposition/non-deposition outcomes in ALD/ASD processes using three-dimensional structural embeddings and process information. A curated dataset of 560 literature-derived ALD/ASD records was constructed, covering 76 precursors, 27 substrates, 20 reactants, and a broad temperature range of 25–620 °C. By separately defining adsorption and deposition as binary classification tasks, the proposed framework captures the sequentially related but mechanistically distinct nature of ALD precursor adsorption and reactant-driven film formation.

Systematic benchmarking across three embedding architectures and three machine learning classifiers demonstrated that CHGNet embeddings combined with an MLP classifier provided the best overall performance, achieving balanced accuracies of 79.2% for adsorption and 84.8% for deposition. Compared with majority-class and classical one-dimensional descriptor baselines, the CHGNet-MLP model showed substantial improvements in both balanced accuracy and average precision, confirming that three-dimensional structural embeddings provide strong predictive value for describing precursor–substrate–reactant relationships in ALD/ASD systems. Detailed ROC, precision–recall, and confusion matrix analyses further showed that deposition outcomes were predicted more robustly than adsorption outcomes, which may reflect the inclusion of reactant information and the more balanced class distribution of the deposition task.

Interpretability analyses revealed chemically meaningful trends in the model behavior. SHAP analysis showed that adsorption predictions are primarily governed by substrate and precursor information, consistent with the surface-mediated nature of precursor adsorption. In contrast, deposition predictions require non-redundant contributions from precursor, substrate, and reactant features, reflecting the coupled roles of adsorbed precursor chemistry, surface environment, and reactant-driven film-forming chemistry. Ablation studies further confirmed that each task-specific input component contributes independently to model performance. In particular, the strong performance drop upon substrate removal highlights the central role of substrate surface chemistry in both adsorption and deposition outcome prediction.

Leave-one-precursor-out validation further demonstrated precursor-dependent transferability beyond simple precursor-specific memorization. For held-out precursors containing both outcome classes, mean balanced accuracies were 66.4% for adsorption and 78.9% for deposition, providing a more conservative estimate of generalization to unseen precursor chemistries. However, the relatively large precursor-to-precursor variance, especially in the adsorption task, indicates that extrapolation to chemically isolated precursor families remains challenging. Representative case studies further showed that the model can distinguish adsorption/non-adsorption and deposition/non-deposition outcomes and can capture reactant- and temperature-dependent deposition behavior in systematic process series. A laboratory-derived experimental comparison further supported this interpretation. When grouped by their relationship to the training data, the model reproduced training-like P0 cases, showed moderate transferability for P1 cases in which the precursor was retained but the substrate was changed, and showed reduced adsorption accuracy for the more chemically diverse P2 cases in which both precursor and substrate differed from the training data. Although deposition predictions remained correct for the limited number of labeled P2 deposition cases, this result should be interpreted together with the small sample size and restricted label availability. These results indicate that AJIN can provide useful prioritization for laboratory screening, while broader chemical-transfer conditions require additional experimentally labeled data for more rigorous validation.

Overall, AJIN provides an experimentally grounded and interpretable screening framework for accelerating ALD process discovery, with potential extensibility toward ASD process design. Rather than replacing experimental validation, the framework is intended to narrow the experimental search space by identifying promising precursor–substrate–reactant–temperature combinations and by highlighting conditions that warrant further investigation. It should be noted that the present framework does not explicitly include several process-specific variables that are important in practical ALD experiments, such as surface termination, precursor dose, exposure time, purge conditions, plasma activation, and incubation-cycle behavior[9] [10] [11]. Similarly, ASD-related predictions should be interpreted as substrate-resolved reactivity trends rather than direct predictions of pattern-level selectivity.

As an initial framework, AJIN is intended to serve as a starting point for broader ALD process prediction, and future versions could incorporate additional experimentally important ALD outcomes beyond adsorption and deposition. The lower specificity observed in the adsorption task further highlights the need to increase the statistical representation of experimentally verified non-adsorption and non-deposition outcomes, particularly for ASD-relevant non-growth conditions. Future work should expand the dataset with systematically curated non-adsorption and non-deposition cases and a broader diversity of precursor, reactant, and substrate chemistries to enable more comprehensive mapping of the ALD/ASD chemical space. In addition, more expressive thermodynamic and kinetic descriptors should be incorporated, particularly to capture non-linear temperature-dependent process windows that are not fully represented by the current scalar temperature input. Extending the framework toward uncertainty-aware prediction will further improve its practical utility for ALD/ASD process design.

## Data Availability

The literature references used to construct the ALD/ASD dataset are provided in Supplementary Data 1.

## Code Availability

The complete codebase, along with usage instructions and example input files, can be accessed at: https://github.com/hbrl-research-group/AJIN-model.

## Acknowledgements

This work was supported by the Korea Collaborative & High-tech Initiative for Prospective Semiconductor Research (K-CHIPS) (2410012107, RS-2025-02311098, and 25074-15FC) and by the Technology Innovation Program (public-private joint investment semiconductor R&D program (K-CHIPS) to foster high-quality human resources) (RS-2023-00236667, High performance Ru-TiN interconnects via high temperature atomic layer deposition (ALD) and development on new interconnect materials based on ALD) funded by the Ministry of Trade, Industry & Energy (MOTIE, Korea; project no. 1415187401).

## Author Contributions

B.G. conceived the study, curated and analyzed the data, developed and implemented the machine-learning framework, performed all computational analyses and validation, prepared the figures, and wrote the original manuscript. T.N.L., Z.M., and M.L. contributed to data preparation and curation. H.-B.-R.L. supervised the study, contributed to the conceptual development of the work, and reviewed and edited the manuscript. All authors reviewed and approved the final manuscript.

## Competing Interests

The authors declare no competing interests.

# Graphical Abstract

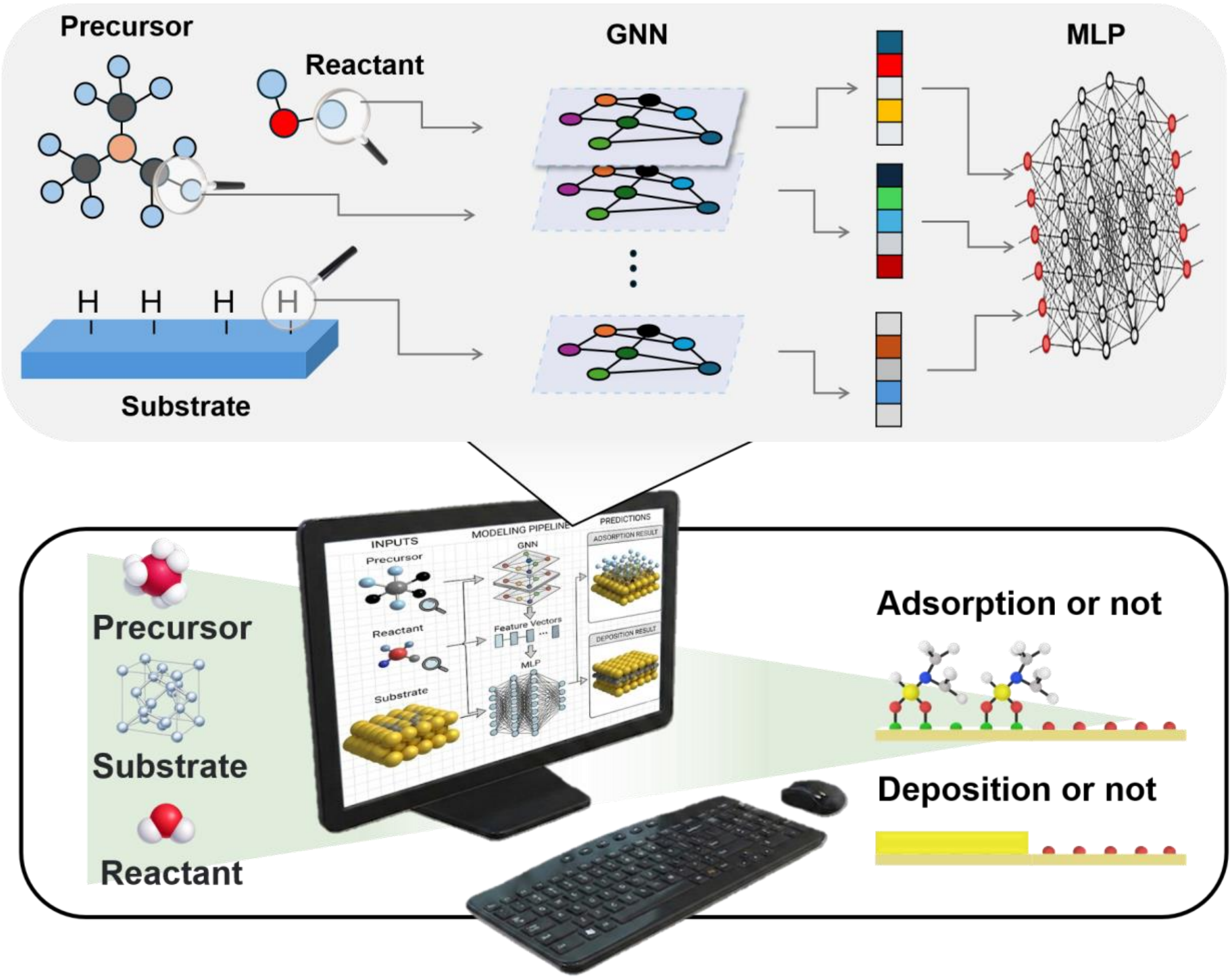

# Supporting

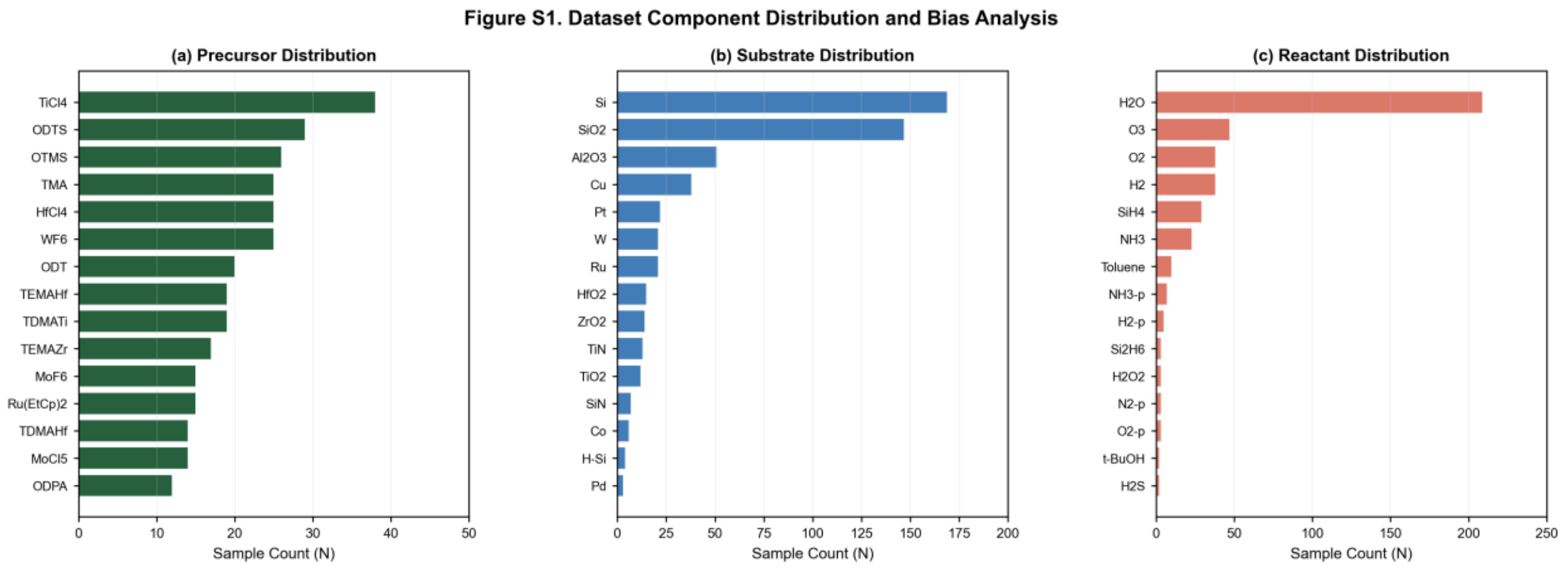

**Figure S1.** Component distribution of the compiled ALD/ASD dataset (N = 560). Sample counts for the (a) precursor, (b) substrate, and (c) reactant components, with each bar representing the number of experimental records containing that species. Plasma-activated reactants are denoted by the suffix "-p" (for example, NH3-p, H2-p).

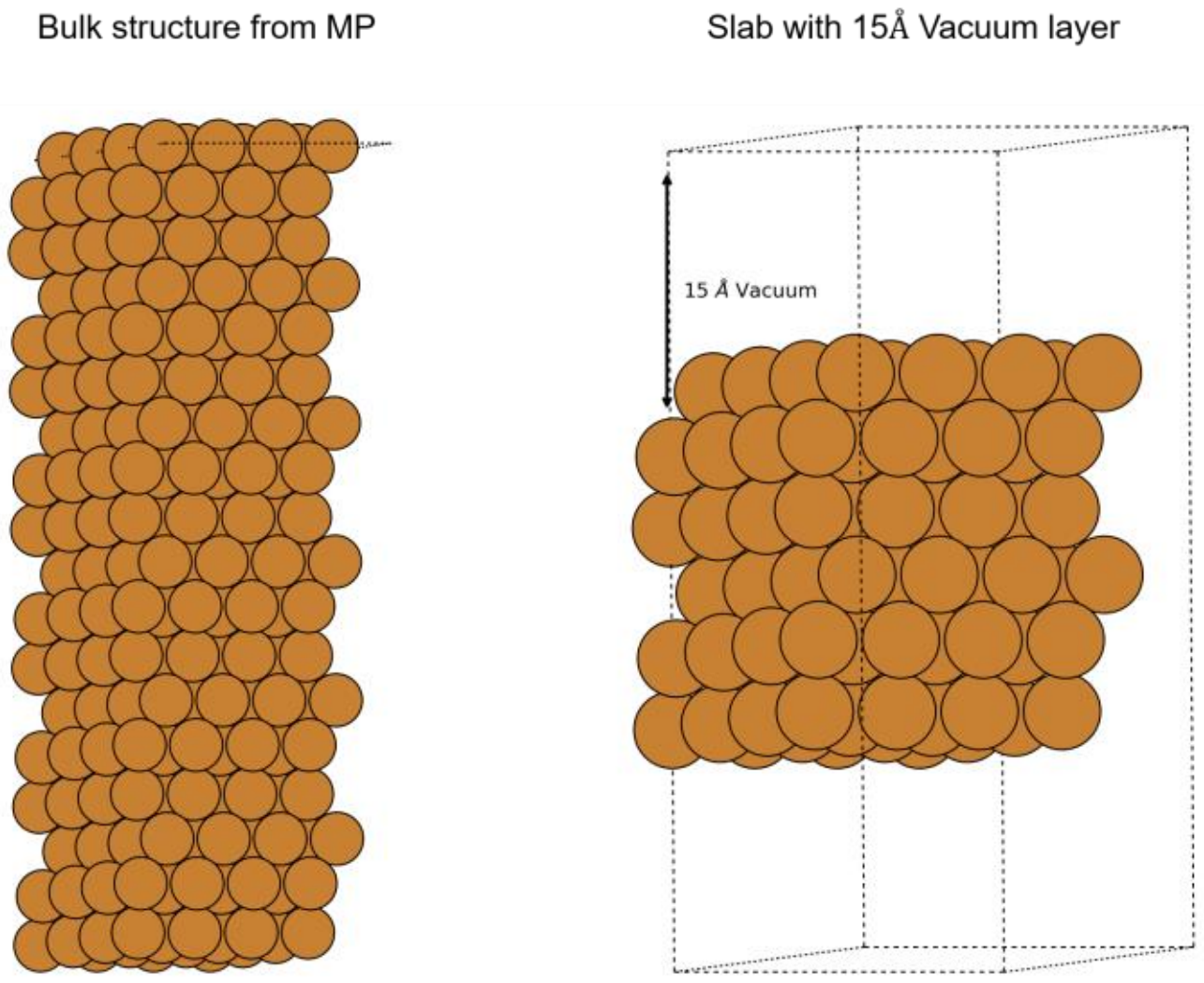

**Figure S2.** (a) The bulk crystal structure (e.g., Cu) obtained from the Materials Project database. (b) The corresponding vacuum-separated bulk-derived representation generated by introducing a 15 Å vacuum region along one cell direction. No explicit crystallographic facet or surface termination was selected. This representation was used to generate a consistent surface-like structural fingerprint for embedding generation.

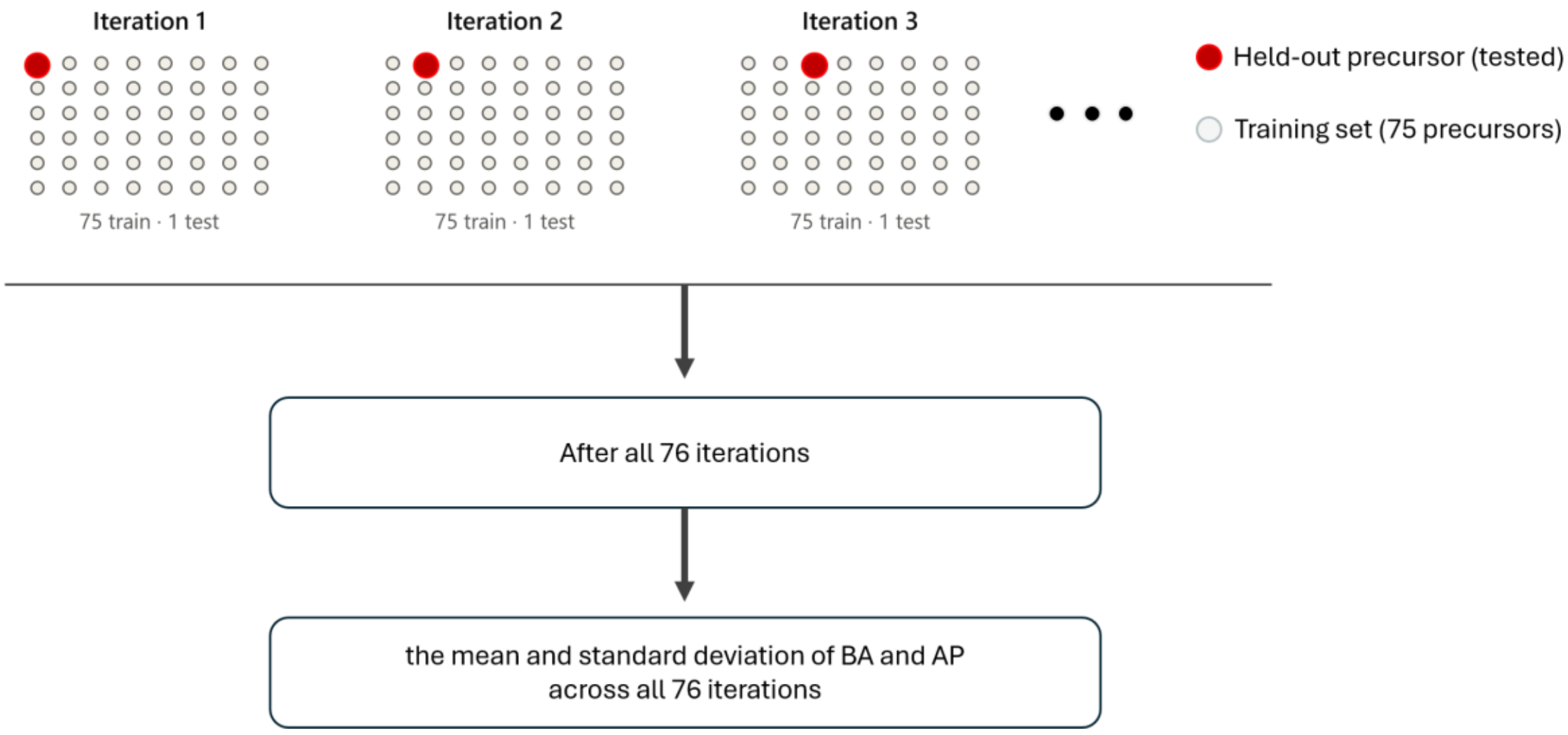

**Figure S3.** Schematic illustration of leave-one-precursor-out (LOPO) cross-validation. In each of 76 iterations, one precursor (red) is held out as the test set while the model is retrained from scratch on the remaining 75 precursors (gray); three representative iterations are shown. After all 76 iterations, BA is summarized across all held-out precursors, whereas AP is summarized only for precursors containing both outcome classes. BA for the same two-class subsets is additionally reported to provide a more conservative assessment of generalization

to unseen precursor chemistries.

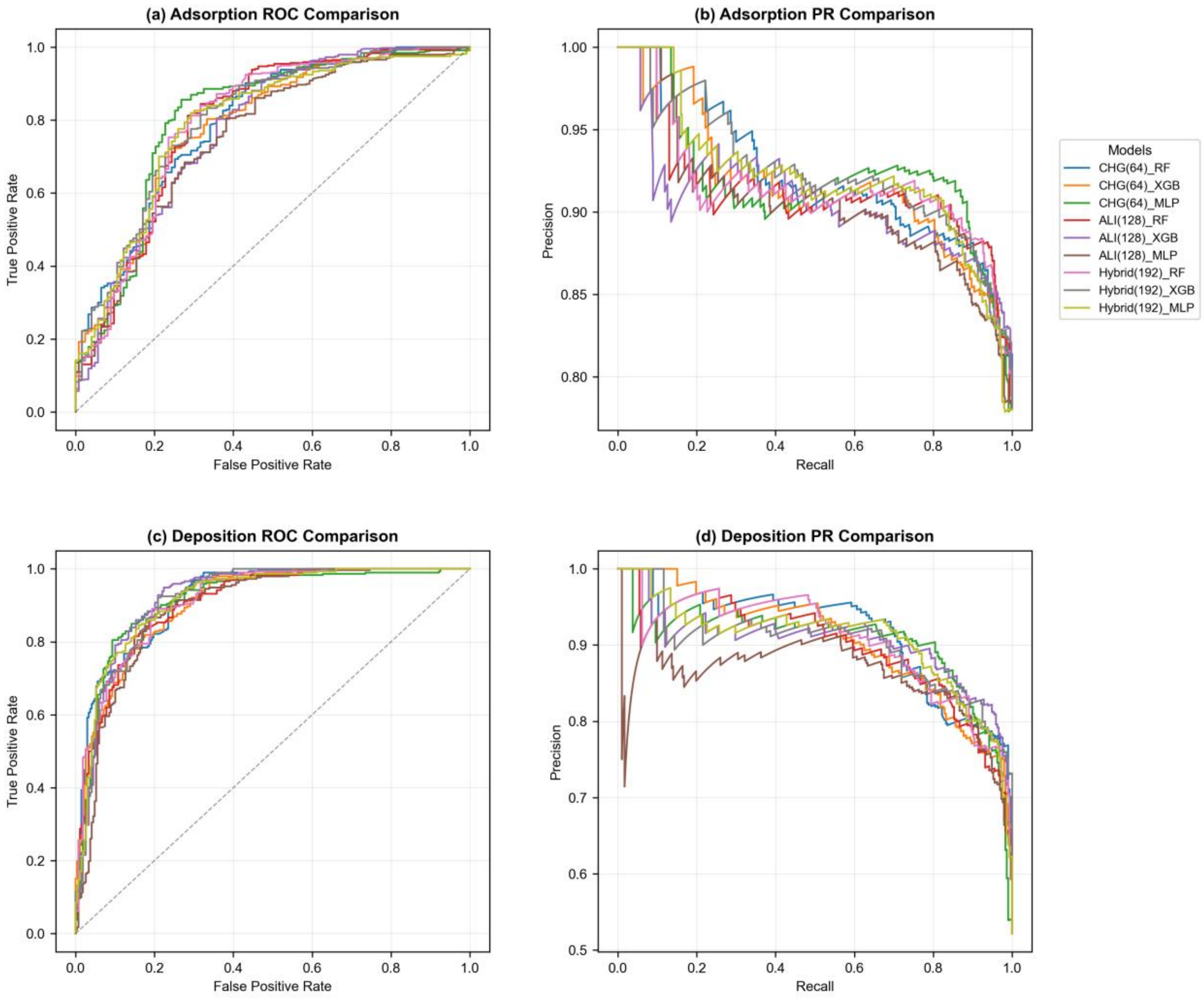

**Figure S4.** ROC curves for (a) adsorption and (b) deposition tasks and precision-recall (PR) curves for (c) adsorption and (d) deposition tasks with all nine models across three GNN embedding architectures (CHGNet, ALIGNN, and combined) and three classifiers (RF, XGB, and MLP).

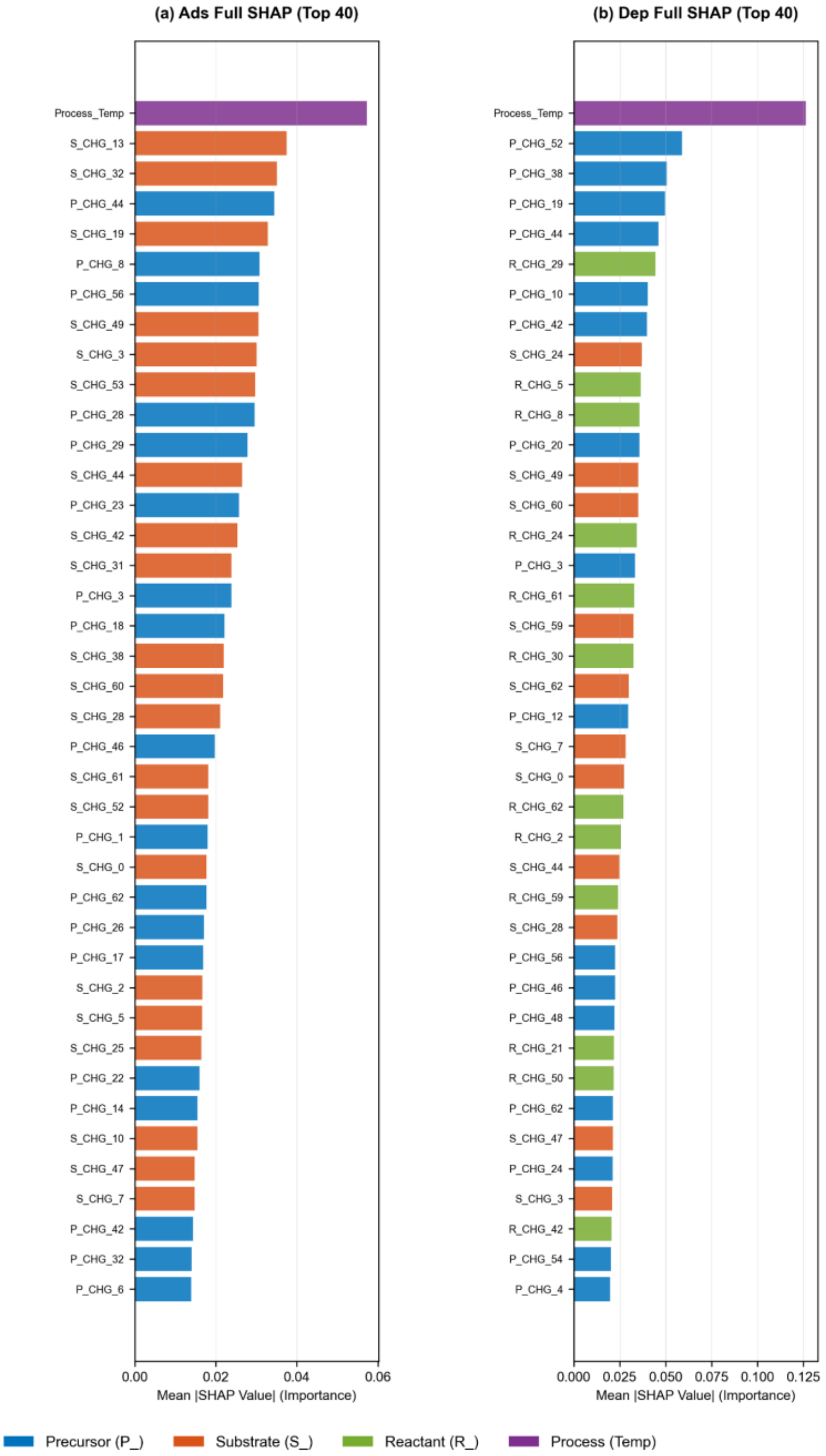

**Figure S5.** SHAP value distributions of the top 40 individual embedding dimensions ranked by mean absolute SHAP value for (a) adsorption and (b) deposition tasks. Feature colors indicate input component type: temperature (purple), precursor (blue), substrate (orange), and reactant (green).

| Model | Hyperparameter | Value | Description |
| --- | --- | --- | --- |
| **Multi-Layer Perceptron (MLP)** | hidden_layer_sizes | (128, 64) | Two hidden layers with 128 and 64 neurons |
| | max_iter | 500 | Maximum number of epochs for optimization |
| | solver | adam | Stochastic gradient-based optimizer |
| | activation | relu | Rectified Linear Unit activation function |
| | batch_size | auto | Minibatch size (approximately 200 samples) |
| | learning_rate_init | 0.001 | Initial learning rate for the optimizer |
| | random_state | 42 | Seed for reproducibility |
| **Random Forest (RF)** | n_estimators | 300 | Number of decision trees in the ensemble |
| | min_samples_leaf | 2 | Minimum samples required at a leaf node to prevent overfitting |
| | class_weight | 'balanced' | Adjusts weights inversely proportional to class frequencies |
| | random_state | 42 | Seed for reproducibility |
| **XGBoost (XGB)** | n_estimators | 300 | Number of gradient boosted trees |
| | learning_rate | 0.05 | Step size shrinkage to prevent overfitting |
| | max_depth | 6 | Maximum depth of a tree to control complexity |
| | eval_metric | 'logloss' | Evaluation metric for validation data |
| | random_state | 42 | Seed for reproducibility |

**Table S1.** Hyperparameter configurations for the three machine learning classifiers (RF, XGB, and MLP) used in this study.

| Feature Name | Category | Description |
| --- | --- | --- |
| rg | Morphological & Geometric Properties | Basic spatial properties of the molecule (e.g., rg) |
| dmax | Morphological & Geometric Properties | Basic spatial properties of the molecule (e.g., dmax) |
| hull_v | Morphological & Geometric Properties | Basic spatial properties of the molecule (e.g., hull_v) |
| hull_a | Morphological & Geometric Properties | Basic spatial properties of the molecule (e.g., hull_a) |
| den | Morphological & Geometric Properties | Basic spatial properties of the molecule (e.g., den) |
| vol | Morphological & Geometric Properties | Basic spatial properties of the molecule (e.g., vol) |
| pack | Morphological & Geometric Properties | Basic spatial properties of the molecule (e.g., pack) |
| planarity | Morphological & Geometric Properties | Basic spatial properties of the molecule (e.g., planarity) |
| chi_m | Basic Physico-chemical Properties | Statistical values (chi) |
| chi_s | Basic Physico-chemical Properties | Statistical values (chi) |
| chi_r | Basic Physico-chemical Properties | Statistical values (chi) |
| rad_m | Basic Physico-chemical Properties | Statistical values (rad) |
| rad_s | Basic Physico-chemical Properties | Statistical values (rad) |
| z_m | Basic Physico-chemical Properties | Statistical values (z) |
| z_s | Basic Physico-chemical Properties | Statistical values (z) |
| std_x, std_y, std_z | Coordinate Statistics | Standard deviation of atomic coordinates along principal axes (x, y, z) |
| rdf_0 – rdf_19 | Radial Distribution Function (RDF) | Structural distribution of interatomic distances across 20 spatial bins (Bins 0 to 19) |
| frac_X (X = H, C, N, O, ...) | Elemental Fractions (Stoichiometry) | Fractional composition of specific elements present in the precursor (H, C, N, O, Al, Si, Ti, Zn, Zr, Hf, Ta, Cl, F, Br, I) |
| extra_0 – extra_29 | Structural One-hot Encodings | Categorical structural identifiers and supplementary topological features (Indices 0 to 29) |

**Table S2.** Classical one-dimensional (1D) molecular descriptors used as baseline features, computed using RDKit. Descriptors are grouped into six categories: morphological and geometric properties, basic physicochemical properties, coordinate statistics, radial distribution function (RDF) features, elemental fractions, and structural one-hot encodings.

**Supplementary Data 1.** Literature references used for dataset construction (Excel file).